\documentstyle[12pt]{article}
\begin{document}
\def\beq{\begin{equation}}
\def\eeq{\end{equation}}
\def\bea{\begin{eqnarray}}
\def\eea{\end{eqnarray}}
\def\ve{\vert}
\def\vel{\left|}
\def\ver{\right|}
\def\nnb{\nonumber}
\def\ga{\left(}
\def\dr{\right)}
\def\aga{\left\{}
\def\adr{\right\}}
\def\rar{\rightarrow}
\def\nnb{\nonumber}
\def\la{\langle}
\def\ra{\rangle}
\def\ba{\begin{array}}
\def\ea{\end{array}}
\def\tep{$B \rar K \ell^+ \ell^-$}
\def\tepm{$B \rar K \mu^+ \mu^-$}
\def\tept{$B \rar K \tau^+ \tau^-$}
\def\ds{\displaystyle}

\def\bos{\lower 0.5cm\hbox{{\vrule width 0pt height 1.3cm}}}
\def\aaa{\lower 0.cm\hbox{{\vrule width 0pt height .8cm}}}
\def\dol{\lower 0.6cm\hbox{{\vrule width 0pt height .8cm}}}

\title{ {\small {\bf TWO HIGGS DOUBLET MODEL AND LEPTON POLARIZATION IN
THE \tept DECAY} } }

\author{\vspace{1cm}\\
{\small T. M. AL\.{I}EV \thanks
{e-mail: taliev@rorqual.cc.metu.edu.tr}\,\,,
M. SAVCI \thanks
{e-mail: savci@rorqual.cc.metu.edu.tr}\,\,,
A. \"{O}ZP\.{I}NEC\.{I} } \\
{\small Physics Department, Middle East Technical University} \\
{\small 06531 Ankara, Turkey} \\
\vspace{5mm}\\  
{\small H. KORU}\\
{\small Physics Department, Gazi University} \\
{\small 06460 Ankara, Turkey} } 

\date{}

\begin{titlepage}
\maketitle
\thispagestyle{empty}

\begin{abstract}
\baselineskip  0.7cm

The decay width, forward-backward asymmetry and $\tau$ lepton
longitudinal and transversal polarization for the exclusive \tept 
decay in a two Higgs doublet model are computed. It is shown that 
the forward-backward asymmetry and longitudinal polarization of 
the $\tau$ lepton are very effective tools for establishing new 
physics.
\end{abstract}

\vspace{1cm}
\end{titlepage}

\section{Introduction}

The analysis of flavor changing neutral current (FCNC) decays is one of the
most promising directions in particle physics, theoretical as well as
experimental, as a potential testing ground for the Standard model (SM) and 
as regards to the physicists' endeavor to comply fully for establishing new 
physics beyond the SM \cite{R3}. Along these lines, the rare $B$ meson 
decays which takes place via the FCNC, play an exceptional role. For example, 
an investigation of these rare decays opens the way for the possibility of a 
more precise determination of the Cabibbo-Kobayashi-Maskawa (CKM) matrix 
elements \cite{R1}. 

Currently the main interest on the rare meson decays is focused on the
decays for which the SM predicts the largest branching ratios that can be
measurable in the near future. The rare \tep ($\ell = e,~\mu,~\tau$)
processes are such decays. For these decays the experimental situation is
quite promising with $e^+~e^-$ and hadron colliders focusing only on the
observation of exclusive modes with lepton pairs as the final states. 
The \tep decay, which is described by $b \rar s \ell^+ \ell^-$
transition at quark level, has been investigated extensively in both SM 
and two Higgs doublet model (2HDM) \cite{R4}-\cite{R17}. 
It is well known that in the 2HDM, the up type quarks acquire their masses
from Yukawa couplings to the Higgs doublet $H_2$ (with the vacuum expectation
value $v_2$) and down type quarks and leptons acquire their masses from Yukawa 
couplings to the other Higgs doublet 
$H_1$ (with the expectation value $v_1$).  In 2HDM there exist five physical
Higgs fields: neutral scalar $H^0$, $h^0$, neutral pseudoscalar $A^0$ and
charged Higgs bosons $H^\pm$. Such a model occurs as natural a feature of
the supersymmetric models \cite{R2}. In these models the interaction
vertex of the Higgs boson and fermions depends on the ratio 
$tan \beta = \frac{\ds{v_2}}{\ds{v_1}}$ which is
a free parameter in the model. The constraints on $tan \beta$ are usually
obtained from $B-\bar B,~ K-\bar K$ mixing, $b \rar s \gamma$ decay width,
semileptonic decay $b \rar c \tau \bar \nu_\tau$ and is given by
\cite{R22,R20}:
\beq
0.7 \le tan \beta \le 0.6 \left( \frac{m_{H^+}}{1~GeV} \right)~,
\eeq
(the lower bound $m_{H^+} \ge 200~GeV$ is obtained in \cite{R20}).

In all these studies the 
contributions from neutral Higgs boson exchange diagrams are neglected, 
since the lepton-lepton-Higgs vertices are proportional to the lepton mass. 
But for the $b \rar s \tau^+ \tau^-$ decay the mass of the $\tau$ lepton is not too
small compared to the $b$ quark mass, and hence one expects that the neutral
Higgs boson exchange diagrams may contribute considerably to such channels.    
It has been pointed by Hewett \cite{R18} that the longitudinal polarization
$P_L$ of the final lepton is an important observable that may be accessible in the
\tept decay mode. Recently it has been shown in \cite{R19} that the
complementary information is contained in $P_L$, together with the two other
orthogonal components of polarization ($P_T$ is the component of of the
polarization lying in the decay plane and $P_N$ is the one that is normal to
the decay  plane). Both $P_T$ and $P_N$ are crucial for the $\tau^+~\tau^-$
channel since they are proportional to
$\frac{\displaystyle{m_\ell}}{\displaystyle{m_b}}$. 
The $b \rar s \ell^+ \ell^-$ transition contains three Wilson coefficients
$C_7,~C_9^{eff}$ and $C_{10}$ in the SM.
The different components
of the polarization, i.e., $P_L$, $P_T$ and $P_N$, involve different
combinations of Wilson coefficients $C_7,~C_9^{eff}$ and $C_{10}$ (see
below) and hence contain independent information. For this reason the
polarization effects are thought to play an important role in further
investigations of the structure of the SM and for establishing new physics beyond
it.

The query for the calculation of the branching ratios and other
observables requires the computation of the   
matrix element of the effective Hamiltonian
responsible for the \tept decay between $B$ and $K$ states.
This problem is related to the non-perturbative sector   
of QCD and it can be solved only by means of a non-perturbative approach.

These matrix elements have been investigated in the framework of different
approaches such as chiral theory \cite{R23}, three point QCD sum rules    
method \cite{R24}, relativistic quark model by the light-front formalism  
\cite{R25}, effective heavy quark theory \cite{R26} and light cone QCD    
sum rules \cite{R27}. The aim of the present work is to calculate these   
matrix elements in the light cone QCD sum rules in the framework of the
2HDM, taking into account the newly appearing operators, $C_{Q_i}$, and to study
the forward-backward asymmetry and final lepton polarization for the 
exclusive \tept decay. Taking into account the additional neutral Higgs
boson exchange diagrams, the effective Hamiltonian is calculated 
in \cite{R28} as  
\bea
{\cal H}_{eff} = \frac{4 G_F}{\sqrt 2} V_{tb} V^*_{ts} \left\{\sum_{i=1}^{10}
C_i( \mu ) O_i( \mu ) + \sum_{i=1}^{10} C_{Q_i}( \mu ) Q_i( \mu ) \right\} ~, 
\eea
where the first set of operators in the curly brackets describe the
effective Hamiltonian responsible for the $b \rar s l^+ l^-$ decay in the SM.
Note that the contributions arising from the diagrams containing the charged Higgs
bosons are taken into account by modifying  the corresponding Wilson
coefficients. These diagrams do not induce any additional operators.
Their explicit forms and the corresponding Wilson coefficients $C_i$ can be 
found in \cite{R7}. The second set of operators in the brackets, whose 
explicit forms are presented in \cite{R28}, come from the exchange of the 
neutral Higgs bosons.
The corresponding Wilson coefficients are:

\bea
C_{Q_1} ( m_W ) &=& - ~\frac{m_b m_\ell}{m_{h^0}^2} tan^2 \beta
\frac{1}{sin^2  \theta_W} \frac{x}{4} \Bigg\{ \left(sin^2  \alpha + 
h\, cos^2  \alpha \right) f_1 (x,y) + \nnb \\
&+& \left[ \frac{m_{h^0}^2}{m_W^2} + \left(sin^2  \alpha + h\,
cos^2  \alpha \right)(1-z) \right] f_2(x,y) +  \nnb \\
&+& \frac{sin^2 2  \alpha}{2 m_{H^\pm}^2} \left[m_{h^0}^2 -
\frac{(m_{h^0}^2 + m_{H^0}^2)^2}{2 m_{H^0}^2} \right] f_3 (y) \Bigg\}
\label{cq1} ~,\\
C_{Q_2} (m_W) &=& \frac{m_b m_\ell}{m_{A^0}^2} tan^2 \beta \left\{ f_1(x,y) +
\left[1+ \frac{m_{H^\pm}^2 - m_{A^0}^2}{m_W^2} \right] f_2(x,y) \right\}
\label{cq2}~,\\
C_{Q_3} (m_W) &=& \frac{m_b e^2}{m_\ell g^2} \Bigg[ C_{Q_1} (m_W) + C_{Q_2}
(m_W) \Bigg] \label{cq3} ~,\\
C_{Q_4} (m_W) &=& \frac{m_b e^2}{m_\ell g^2} \Bigg[ C_{Q_1} (m_W) -
C_{Q_2}(m_W) \Bigg] \label{cq4}~, \\ 
C_{Q_i}(m_W) &=& 0 ~~~~~~ i =5, \ldots,10 \label{cqi}~,
\eea
where
\newpage
$$x = \frac{m_t^2}{m_W^2}~,~~~~y=\frac{m_t^2}{m_{H^\pm}^2}~,~~~~z=
\frac{x}{y}~,~~~~h=\frac{m_{h^0}^2}{m_{H^0}^2}~, $$
$$f_1 (x,y) = \frac{x\, lnx}{x-1} - \frac{y\, lny}{y-1}~,~~~~ f_2(x,y) =
\frac{x\, lny}{(z-x)(x-1)} + \frac{lnz}{(z-1)(x-1)}~,$$
$$f_3(y) = \frac{1 -y + y\, lny}{(y-1)^2}~.$$

The QCD correction to the Wilson coefficients $C_i(m_W)$ and $C_{Q_i}(m_W)$ can be
calculated using the renormalization group equations.
In \cite{R28} it was shown that the operators $O_9$ and $O_{10}$ do not
mix with $Q_i~(i=1, \ldots,10)$, so that the Wilson coefficients $C_9$ and
$C_{10}$ remain unchanged and their values are the same as in the SM. Their explicit
forms can be found in \cite{R28}, where it is also shown
that $O_7$ can mix with $Q_i$.
But additional terms due to this mixing
can  safely be neglected since the corrections to the
SM value of $C_7$ arising from these terms are less than $5 \%$ when $tan \beta \le 50$.

Moreover the operators $O_i~(i=1, \ldots,10)$ and $Q_i~(i=3, \ldots,10)$ do
not mix with $Q_1$ and $Q_2$ and also there is no mixing between $Q_1$ and
$Q_2$. For this reason the evolutions of the coefficients $C_{Q_1}$ and
$C_{Q_2}$ are controlled by the anomalous dimensions of $Q_1$ and $Q_2$
respectively:
$$C_{Q_i} (m_b) = \eta^{-\gamma_Q / \beta_0} C_{Q_i} (m_W)~,~~~i=1,~2 ,$$
where $\gamma_Q = -4 $ is the anomalous dimension of the operator $\bar s_L b_R$.

Neglecting the strange quark mass, the matrix element for $b \rar s \tau^+
\tau^-$ decay is \cite{R28}:
\bea
{\cal M} &=& \frac{G_F \alpha}{2\sqrt 2 \pi} V_{tb} V^*_{ts} \Bigg\{C_9^{eff}
\bar s \gamma_\mu (1- \gamma_5) b \, \bar \tau \gamma^\mu \tau + C_{10} \bar s
\gamma_\mu (1- \gamma_5) b \, \bar \tau \gamma^\mu \gamma_5 \tau -  \nnb \\
&-& 2C_7\frac{m_b}{p^2}\bar s i \sigma_{\mu \nu}p^\nu (1+\gamma_5)  b  \,  
\bar \tau \gamma^\mu \tau + C_{Q_1} \bar s (1 + \gamma_5) b \bar \tau \tau +
C_{Q_2} \bar s (1+\gamma_5) b  \bar \tau \gamma_5 \tau \Bigg\}~, \label{m}
\eea
where $p^2$ is the invariant dileptonic mass, the Wilson coefficients $C_7,~C_9$ 
and $C_{10}$ are obtained from their SM values by adding the contributions
due to the charged Higgs bosons exchange diagrams. Note that this addition
is performed at high $m_W$ scale, and then using the renormalization group
equations, the coefficients are calculated at lower  $m_b$ scale.
Coefficients $C_{Q_1}$ and
$C_{Q_2}$ describe the neutral Higgs boson exchange diagrams' contributions. 
Note that the coefficient $C_9^{eff}(\mu,p^2) \equiv C_9( \mu) + Y ( \mu,~p^2)$, 
where the function $Y$ contains the contributions from the one loop matrix element
of the \cite{R7,R30,R31}. In addition to the short distance contributions, 
it is possible to take into account the long distance effects associated with real 
$c\bar c$ in the intermediate states, i.e., with the cascade process 
$B \rar K J/\psi(\psi^\prime)\rar K \ell^+ \ell^-$.These contributions
are taken into account by introducing a Breit-Wigner  
form of the resonance propagator and this procedure leads to an additional
contribution to $C_9^{eff}$ of the form \cite{R11,R32}
\bea
-\, \frac{3 \pi}{\alpha^2} \sum_{V= J/\psi,~ \psi^\prime,\ldots} 
\frac{m_V \Gamma(V \rar \ell^+ \ell^-)}{(p^2 - m_V^2) - i m_V \Gamma_V}~. \nnb
\eea
From eq.(\ref{m}) it is obvious that, in order to calculate the decay width and
other observables for the exclusive \tep channel, the matrix elements
$ \la K\vel \bar s \gamma_\mu (1- \gamma_5) b \ver B \ra$, 
$\la K \vel \bar s i \sigma_{\mu \nu} q^\nu (1+\gamma_5) b \ver B \ra$,
and $ \la K \vel \bar s (1+\gamma_5) b \ver B \ra$ have to be calculated.
These matrix elements can be parametrized in terms of the formfactors
$f^+,~f^-$ and $f_T$ in the following way:
\bea
\la K \ga p_K \dr \vel \bar s \gamma_{\mu}( 1- \gamma_5) b \ver B \ga p_B \dr \ra &=&
\ga p_B + p_K \dr_\mu f^+ (p^2) + p_\mu f^- ( p^2)~,\label{gm1g5} \\ \nnb \\
\la K \ga p_K \dr \vel \bar s i \sigma_{\mu \nu} p^\nu (1+ \gamma_5) b
\ver B \ga p_B \dr \ra &=&
\Big[ \ga p_B + p_K \dr_\mu p^2 - \nnb \\
&-& p_\mu (m_B^2 - m_K^2 ) \Big] 
\frac{f_T ( p^2 )}{m_B+m_K}~,
 \label{si1g5}
\eea
where $p=p_B - p_K$ is the momentum transfer. To be able to calculate the the matrix
element $\la K \vel \bar s (1+ \gamma_5) b \ver B \ra$, we multiply both
sides of eq.(\ref{si1g5}) by $p_\mu$ and use the equation of motion. Neglecting the mass
of the strange quark, we get:
\bea
\la K(p_K) \vel \bar s (1+ \gamma_5) b \ver B(p_B) \ra &=&
\Big[(m_B^2-m_K^2) f^+(p^2)+p^2 f^-(p^2) \Big] \frac{1}{m_b} \label{1g5} ~. 
\eea
Making use of eqs.(\ref{gm1g5}), (\ref{si1g5}) and  (\ref{1g5}) we obtain for the matrix element
of the \tept decay:
\bea
{\cal M} &=& \frac{G \alpha}{2 \sqrt 2 \pi} V_{tb} V_{ts}^*  
\Bigg\{ \Big[ A {p_K}_\mu + B p_\mu \Big] \bar \ell \gamma^\mu \ell 
+ \Big[ C {p_K}_\mu + D p_\mu \Big] \bar \ell \gamma^\mu \gamma_5 \ell
+ F_1 \bar \ell \ell + F_2 \bar \ell \gamma_5 \ell \Bigg\}~, \label{m2}
\eea
where 
\bea
A&=& 2\, C_9^{eff}f^+ - C_7\frac{4 m_b f_T(p^2)}{m_B+m_K}~,\nnb\\
B&=& C_9^{eff}\left[ f^-(p^2) + f^+(p^2) \right] +
C_7\frac{2 m_b f_T(p^2)}{p^2} \frac{(m_B^2 - m_K^2 - p^2)}{m_B + m_K}~, \nnb \\
C&=& 2\, C_{10}f^+(p^2) ~, \nnb \\
D&=& C_{10} \left[f^-(p^2) + f^+(p^2) \right] ~,\nnb \\
F_1&=& C_{Q_1} \frac{1}{m_b} \Big[ (m_B^2-m_K^2) f^+(p^2) + p^2 f^-(p^2)
\Big]~, \nnb \\
F_2&=& C_{Q_2} \frac{1}{m_b} \Big[ (m_B^2-m_K^2) f^+(p^2) + p^2
f^-(p^2)\Big]~.
\eea
The formfactors $f^+(p^2)$, $f^-(p^2)$ and $f_T(p^2)$ 
are investigated in the light cone QCD sum rules framework and to a good
accuracy their $p^2$ dependence are found to be representable in the
following pole forms \cite{R27}:
\bea
f^+ (p^2) &=& \frac{0.29}{\left(1 - \displaystyle{\frac{p^2}{23.7}}\right)}~, \nnb \\
f^-(p^2) &=& - \frac{0.21}{\left(1 - \displaystyle{\frac{p^2}{24.3}}\right)}~, \nnb \\
f_T(p^2) &=& - \frac{0.31}{\left(1 - \displaystyle{\frac{p^2}{23}}\right)}~,\label{pfm}
\eea
which we will use in the numerical calculations. Using eq.(\ref{m2}) and
performing summation over final  lepton polarization, we get for the double
differential decay rate:
\bea
\frac{d \Gamma}{dp^2 dz} &=& \frac{G^2 \alpha^2 }{2^{12} \pi^5} \, \frac{
\vel V_{tb} V_{ts}^* \ver^2 v \sqrt \lambda}{m_B} \Bigg\{ \frac{1}{2} \lambda m_B^4
\vel A \ver ^2 + \frac{1}{2} \vel C \ver^2  m_B^2 \left(\lambda m_B^2
+ 16 m_\ell^2 r \right) + 2 \vel F_2 \ver^2 m_B^2 s + \nnb \\
&+&  8 Re(D^*F_2) m_B^2 m_\ell s + 8 \vel D
\ver ^2 m_B^2 m_\ell^2 s + 4 Re(C^* F_2) m_B^2 m_\ell \left( 1-r-s \right) + \nnb \\
&+& 8 Re (C^* D) m_B^2 m_\ell^2(1-r-s) + 2 \vel F_1 \ver^2 m_B^2 s v^2 + 
z \left[ 4 Re( A^* F_1) \sqrt{\lambda} m_B^2 m_\ell v \right] - \nnb \\
&-&  \frac{z^2}{2} \lambda m_B^4  v^2 (\vel A \ver ^2 + \vel C \ver ^2)
\Bigg\}~, \label{ddr}
\eea
where $z=cos  \theta$ and $\theta$ is the angle between the three momenta of
the negatively charged lepton and the $B$-meson in the CM frame of the final
leptons, and $v=\sqrt{{1-\frac{4 m_\ell^2}{p^2}}}$ is the lepton velocity. Here
$\lambda (1,r,s)$ is the usual triangle function and
$r=\frac{m_K^2}{m_B^2},~s=\frac{p^2}{m_B^2}$.
As we have noted previously, the forward-backward asymmetry $A_{FB}$ and the
final lepton polarization involve different combinations of the Wilson
coefficients $C_7,~C_9^{eff},~C_{10},~C_{Q_1}$, and $C_{Q_2}$ and therefore
each of them contains independent information. For this reason, here in what
follows we study these quantities in more detail.

The forward-backward asymmetry $A_{FB}$ is defined as:
$$ A_{FB} (p^2) = \frac{\displaystyle{\int_0^1 dz \frac{d \Gamma}{dp^2 dz} -
\int_{-1}^0dz
\frac{d \Gamma}{dp^2 dz}}}{\displaystyle{\int_0^1 dz \frac{d \Gamma}{dp^2
dz}+
\int_{-1}^0dz\frac{d \Gamma}{dp^2 dz}}}~.$$

Note that in the SM, the forward-backward asymmetry is zero when the
polarization of the final lepton is summed over. The reason is obvious: the
hadronic current for the $B \rar K$ transition is a pure vector current. But
forward backward asymmetry (or charge asymmetry) is non-zero only if there
exists C-violating terms. In the 2HDM there is a C-violating term proportional
to $F_1$ (see eq.(\ref{m2})), so that, $A_{FB}$ is nonzero
and is proportional to the lepton mass. For small values of $tan \beta$, 
the contributions from the neutral Higgs boson exchange diagrams are
very small, and hence one expects that the value of $A_{FB}$ be small also.
But for large $tan \beta$, the contributions of the neural Higgs boson
exchange diagrams become significant and we  expect $A_{FB}$ to be
large. The numerical analysis confirms these expectations (see numerical
analysis section).

Let us now discuss the lepton polarization effects. We define three
orthogonal unit vectors:
\bea
\vec{e}_L &=& \frac{\vec{p}_1}{\vel \vec{p}_1 \ver}~, \nnb \\
\vec{e}_N&=& \frac{\vec{p}_{K} \times \vec{p}_1}
{\vel \vec{p}_{K} \times \vec{p}_1 \ver}~, \nnb \\
\vec{e}_T &=& \vec{e}_N \times \vec{e}_L~, \nnb
\eea
where $\vec{p}_1$ and $\vec{p}_{K}$ are the three momenta of the
$\ell^-$ lepton
and the $K$ meson, respectively, in the center of mass of the
$\ell^+~\ell^-$ system. The differential decay rate for any given spin
direction $\vec{n}$ of the $\ell^-$ lepton, where $\vec{n}$ is a unit vector  
in the $\ell^-$ lepton rest frame, can be written as
\bea
\frac{d \Gamma \ga \vec{n} \dr}{d p^2} = 
\frac{1}{2} \ga \frac{d \Gamma}{d p^2} \dr_{\!\!\! 0} 
\Big[ 1 + \ga P_L\, \vec{e}_L + P_N\, \vec{e}_N + P_T\, \vec{e}_T \dr \cdot
\vec{n} \Big]~,
\eea
where the subscript "0" corresponds to the unpolarized case, and $P_L,~P_T$,
and $P_N$, which correspond to the longitudinal, transverse and normal
components of the polarization vector, respectively, are functions of $p^2$.
These  components $P_i~(i=L,~T,~N)$ are defined as:
\bea
P_i (p^2) = \frac{ {\displaystyle{\frac{d \Gamma}{dp^2}
\ga \vec{n}=\vec{e}_i \dr -
\frac{d \Gamma}{dp^2}\ga \vec{n}=-\vec{e}_i \dr}} }
{ {\displaystyle{\frac{d \Gamma}{dp^2}\ga \vec{n}=\vec{e}_i \dr +
\frac{d \Gamma}{dp^2}\ga \vec{n}=-\vec{e}_i \dr}} } ~.
\eea

The calculations for the $P_i$'s (i = $L,~T$) lead to the following results:
\bea
P_L &=& \frac{v}{\Delta} \Big[ \frac{2}{3} \lambda m_B^4 Re( A^* C) - 
4 Re (F_1^* F_2) m_B^2 s - 8 Re (D^* F_1) m_B^2 m_\ell s - \nnb \\
&-& 4 Re (C^* F_1) m_B^2 m_\ell (1-r-s) \Big]  \label{eq1}~, \\ \nnb \\
P_T &=& \frac{\pi \sqrt{\lambda} m_B^3}{ {\sqrt s} \Delta} \Big[
m_\ell (1-r-s) Re (A^* C) + s v^2 
Re (C^* F_1) +\nnb \\
&+& s  Re (A^* F_2)  + 2 s m_\ell Re (A^* D)  \Big]~. \label{eq2}
\eea
The factor $\Delta$ in eqs. (\ref{eq1}) and  (\ref{eq2})  can be obtained
from eq.(\ref{ddr}) by an integration over $z$ of the terms in the curly
brackets. Note that the explicit form of the normal component $P_N$ of the 
polarization vector of the $\ell^-$ lepton is also calculated. However, an 
analysis of its behavior with respect to $p^2$ shows that numerically 
it is  quite small, so that we do not present it.   
As a check of our results, when we equate $F_1$ and $ F_2$
to zero, i.e., neglect the contributions from the Higgs bosons, we obtain the 
results of \cite{R27}.

\section{Numerical Analysis}
The values of the main input parameters, which appear in the expression for the decay width
are: $m_b=5~GeV,~m_c=1.4~GeV,~m_\tau = 1.78~GeV,~m_\mu = 0.105~GeV,
~\Lambda_{QCD} = 225~ MeV,~m_B = 5.28~GeV$,
and $m_{K}=0.495~GeV$. We use the pole form of the formfactors given in
eq.(\ref{pfm}). For $B$ meson
lifetime we take $\tau(B_d)=1.56 \times 10^{-12}~s$ \cite{R33}. The values of the Wilson
coefficients $C_7^{SM} (m_b)$ and $C_{10}^{SM} (m_b)$ to the leading logarithmic
approximation are \cite{R34,R35}:
$$ C_7 = -0.315~,~~~C_{10} = -4.642~.$$
The expression $C_9^{eff}$ for the $b \rar s$ transition in the next to
leading order approximation is given as (see for example \cite{R34}):
\\ 
\bea
\lefteqn{
C_9^{eff} (m_b) = } \nnb \\
&& C_9^{SM} (m_b) + C_9^{H^-} (m_b) + 0.124 w( \hat{s}) + g( \hat{m_c},\hat{s}) \left( 3 C_1 +
C_2 + 3 C_3 + C_4 + 3 C_5 + C_6 \right) - \nnb \\
&-&\frac{1}{2} g( \hat{m_q},\hat{s})
\left( C_3 + 3 C_4 \right) - \frac{1}{2} g( \hat{m_b},\hat{s}) \left( 4C_3 +
4 C_4 + 3 C_5 + C_6 \right) + \nnb \\
&+& \frac{2}{9} \left( 3 C_3 + C_4 + 3 C_5 + C_6
\right)~,
\eea  
with $$ C_1 = -0.249~,~~ C_2 = 1.108~,~~C_3 = 1.112 \times 10^{-2}~,~~ C_4
= -2.569 \times 10^{-2}~,$$ $$ C_5 = 7.4 \times 10^{-3}~,~~C_6 = -3.144
\times 10^{-2}~,~~C_9^{SM} (m_b) = 4.227,$$ where $\hat{m_q}=\frac{\ds{m_q}}{\ds{m_b}},~ \hat{s} =
\frac{\ds{p^2}}{\ds{m_b^2}}$. The explicit forms of $C_7^{H^-}(m_W)$, 
$C_9^{H^-}(m_W)$ and $C_{10}^{H^-}(m_W)$ can be found in \cite{R6}.

In the above expression $ w(\hat{s})$ represents the one gluon correction to
the matrix element $O_9$  and its explicit
form  can be found in \cite{R13},
while the function $g( \hat{m_q},\hat{s})$ arises from the one loop
contributions of the four quark operators $O_1 - O_6$ (see for example
\cite{R34,R35}), i.e.,
\bea
g(\hat{m_q},\hat{s^\prime}) &=& - \frac{8}{9} ln\, \hat{m_q} + \frac{8}{27} +
\frac{4}{9} y_q - \frac{2}{9} ( 2 + y_q) \sqrt { 11 - y_q} + \nnb \\
&+& \left\{ \theta (1-y_q) \left( ln \frac{1+ \sqrt{1-y_q}}{1-
\sqrt{1-y_q}} - i \pi \right) + \theta(y_q -1) arctan \frac{1}{\sqrt{y_q -1}}
\right\}~, 
\eea
where $y_q = \frac{\ds{\hat{{m_q}}}}{\ds{\hat{s^\prime}}}$, and $\hat{s^\prime} =
\frac{\ds{4 p^2}}{\ds{m_b^2}}$.

In Table 1 we list the 
three different sets of values for the masses of the Higgs particles 
$m_{h^0}$, $m_{H^\pm}$, $m_{H^0}$ and $m_{A^0}$ that we use throughout
the numerical calculations.   


\begin{table}[h]
\begin{center}
\begin{tabular}{|c|c|c|c|c|}
\hline
      &\bos $ m_{h^0} $ & $ m_{H^\pm} $ & $ m_{H^0} $ & $ m_{A^0} $
\\  \hline\hline
mass set-1  & \bos $  80~GeV $ & $ 200~GeV $ & $ 150~GeV $ & $ 100~GeV $ \\
\hline
mass set-2  & \bos $ 250~GeV $ & $ 300~GeV $ & $ 100~GeV $ & $ 350~GeV $ \\
\hline
mass set-3  & \bos $ 100~GeV $ & $ 400~GeV $ & $ 200~GeV $ & $ 150~GeV $ \\
\hline
\end{tabular}
\vskip 0.3 cm
\caption{List of the values for the masses of the Higgs particles.}
\end{center}
\end{table}


Further, we choose $sin  \alpha = \frac{\ds{\sqrt 2}}{\ds{2}}$ and 
the numerical calculations for the branching ratio are all carried 
at three different values of $tan \beta$, i.e.,
$tan \beta = 1,~tan \beta = 20$ and $tan \beta =30$.

In Fig.1 (a)  we present the $p^2$ dependence of the differential 
branching ratio
for the $B_d \rar K \tau^+ \tau^-$ decay with the long distance
effects, for $tan \beta = 1$. The numbers 1, 2 and 3 on each curve
identify the mass set-1, 2 and 3 for the Higgs particles $m_{H^\pm}$,
$m_{h^0}$, $m_{H^0}$ and $m_{A^0}$, respectively, as displayed in Table 1. 
The curve numbered as 4 is the one calculated in the Standard Model
\cite{R27} and is depicted for a comparison of the two models.   
Fig.1 (b) and (c) are the similar graphs at $tan \beta = 20$ and $tan 
\beta =30$, respectively. The sharp peaks in
these figures are due to the long distance contributions. Note that the 
curves that represent the short distance contributions are not plotted in this 
set of figures. The reason for this is our observation of the fact that, 
in each respective mass set
these curves overlap with and mimic the behavior of the 
curves representing the long distance effects at all points of $p^2$, 
except at sharp peaks.
It is also observed that the 
spectrum of the invariant mass distribution is slightly asymmetric.

In Fig.2 (a) and (b) we plot the dependence of the forward-backward asymmetry $A_{FB}$ 
on $p^2$ for the $B_d \rar K \tau^+ \tau^-$ decay,
with and without the long-distance effects at different values of
$tan \beta$. The lines numbered as 1, 3 and 5 represent the long distance
effects for the mass sets 1, 2 and 3, respectively, while the lines with numbers 
2, 4 and 6 are depicted for the short distance effects. 
From these figures we see that $A_{FB}$ is negative for all
values of $p^2$ except in the $\psi^\prime$ resonance region and it is sensitive
to the value of $tan \beta$. For $tan \beta = 1$ $A_{FB}$ is
quite small, so that we do not present it here.
 
In Fig.3 (a), (b) and (c) we present the $p^2$ dependence of the 
longitudinal polarization
of the final lepton $P_L$ without the long distance effects at
$tan \beta = 1$, $tan \beta = 30$ and $tan \beta = 50$,
respectively. The lines numbered as 1, 2 and 3
represent the mass set-1, 2 and 3, respectively. The line with number 4 is
the one presented in \cite{R27} for the Standard Model calculations. 
As is obvious from these figures if we exclude the
resonance mass region of $\psi^\prime$, $P_L$ is negative for all values of
$p^2$. Note that the behavior of the $P_L$ curves are sensitive to the
different choices of the value of $tan \beta$.  

In Fig.4 (a), (b) and (c) we present the $p^2$  dependence of the 
transversal polarization $P_T$ of the $\tau$ lepton which lies 
in the decay plane, without the long distance effects, at 
$tan \beta=1$, $tan \beta=30$ and at  $tan \beta=50$, respectively.
The lines numbered as 1, 2 and 3
represent the mass set-1, 2 and 3, in the respective order. 
From these figures it follows that at $tan \beta=1$  $P_T$ 
is positive at all values of $p^2$. On the other hand, for $tan \beta=30$ and $tan
 \beta=50$, $P_T$ is positive near the threshold region while it becomes
negative far from the threshold region. 
Therefore the determination of the sign of $P_T$ in the future experiments is a
very important issue and can provide a direct information   
for the establishment of new physics.

For completeness in Figs.5-7 we present the results for the \tepm decay. It
is clear that in this case the neutral Higgs exchange diagram contributions
are quite small and the deviation from the SM prediction is due to the
charged Higgs boson exchange diagrams.  

In Fig.5 we present the differential branching ratio versus $p^2$ for \tepm
with and without the long distance effects. The sharp peaks in
these figures are
due to the long distance contributions, as is the case for Fig.1.
Fig.5 (a) is for the mass set-1, (b) is for the mass set-2 and (c) is presented for 
the mass set-3. 
The lines numbered as 1, 3 and 5 represent the long distance
effects for the mass sets 1, 2 and 3, respectively, while the numbers 
2, 4 and 6 are
presented for the short distance effects. The abbreviation $SM$ on the
bottom pair of curves (5 and 6) stands for the Standard Model results 
as depicted in \cite{R27}.  
The branching ratio for the \tepm decay is not as sensitive to the value of   
$tan \beta$ as for the \tept decay. 

The behavior of the longitudinal polarization $P_L$ with changing $p^2$ for the
$B \rar K \mu^+ \mu^-$ decay is presented in Fig.6, 
with and without the long distance effects. 
Curves 1, 2 are evaluated numerically at $tan \beta = 1$ while 
those 3 and 4 represents the behavior of the 
longitudinal polarization  calculated at
$tan \beta = 50$. The last two curves 5 and 6 depict the Standard Model
results \cite{R27}.
The ordering of the figures and of the
numbering of lines for the short and long distance effects 
are exactly the same as explained for Fig.5. Without the 
long distance effects $P_L$ is always negative.

In Fig.7 (a), (b) and (c) we present the transversal polarization $P_T$ for 
the \tepm decay as a
function of $p^2$, for mass set-1, mass set-2 and mass set-3, respectively. 
While at $tan \beta = 1$ $P_T$ is always positive, at $tan
\beta = 30$ and $tan \beta = 50$ it changes sign. Thus the investigation of
the sign of $P_T$ can be an effective tool in search for new physics.

Since, as we have already noted, $A_{FB},~P_L$, and $P_T$ contain
independent information, their investigation in the future experiments will be
quite an efficient tool for establishing new physics.
 
In Table 2 we present the values of the branching ratios
for the $B_d \rar K \tau^+ \tau^-$ decay.
After integrating over $p^2$ we get for the branching ratios for the 
$B_d \rar K \tau^+ \tau^-$ decay, with the long distance contributions:


\begin{table}[h]
\begin{center}
\begin{tabular}{|c|c|c|c|}
\hline
      &\bos ~\,$ tan \beta = 1 $~\, & ~\,$ tan \beta = 20 $~\, & ~\,$
tan \beta = 30 $~\,
\\  \hline\hline
mass set-1  & \bos $ 3.47 \times 10^{-7} $ & $ 2.68 \times 10^{-7} $ 
& $ 4.12 \times 10^{-7} $ \\ \hline
mass set-2  & \bos $ 3.15 \times 10^{-7} $ & $ 2.64 \times 10^{-7} $ 
& $ 3.71 \times 10^{-7} $ \\ \hline
mass set-3  & \bos $ 2.96 \times 10^{-7} $ & $ 3.01 \times 10^{-7} $
& $ 5.43 \times 10^{-7} $ \\ \hline
\end{tabular}
\vskip 0.3 cm
\caption{Branching ratios for the exclusive $B_d \rar K \tau^+ \tau^-$
decay, with the long distance contributions.}
\end{center}
\end{table} 


The ratio of the exclusive and inclusive channels is defined as:
$$R = \frac{\displaystyle{B(B_d\rar K \tau^+ \tau^-)}}
         {\displaystyle{B(b \rar s \tau^+ \tau^-)}}~.$$

\newpage

In the SM this ratio is given as $R = 0.07 \pm 0.02$ when
$B(b \rar s \tau^+ \tau^-) = (2.6 \pm 0.5) \times 10^{-7}$ \cite{R35}.
In Table 3 we display the results that we have calculated for our case, using the
values given in Table 2. 


\begin{table}[h]
\begin{center}
\begin{tabular}{|c|c|c|c|}
\hline
      &\bos ~\,$ tan \beta = 1 $~\, & ~\,$ tan \beta = 20 $~\, & ~\,$
tan \beta = 30 $~\,
\\  \hline\hline
mass set-1  & \bos 1.33 & 1.03 & 1.58  \\ \hline
mass set-2  & \bos 1.21 & 1.02 & 1.43  \\ \hline
mass set-3  & \bos 1.14 & 1.16 & 2.09  \\ \hline
\end{tabular}
\vskip 0.3 cm
\caption{The ratio $R$ of the exclusive $B(B_d\rar K \tau^+ \tau^-)$ 
and SM value of the inclusive $B(b \rar s \tau^+ \tau^-)$ channels.}
\end{center} 
\end{table}

\noindent where we have used the SM value for the inclusive $B(b \rar s \tau^+ \tau^-)$.

For completeness we present in Table 4 the branching ratio with respect to the inclusive 
$B(b \rar s \tau^+ \tau^-)$ decay in 2HDM with the long distance contributions. 
The expression for the inclusive $B(b \rar s \tau^+ \tau^-)$ decay is given in \cite{R28}. 

\begin{table}[h]
\begin{center}
\begin{tabular}{|c|c|c|c|}
\hline
      &\bos ~\,$ tan \beta = 1 $~\, & ~\,$ tan \beta = 20 $~\, & ~\,$
tan \beta = 30 $~\,
\\  \hline\hline
mass set-1  & \bos $ 18.90 \times 10^{-7} $  & $ 17.97 \times 10^{-7} $ 
& $ 18.44 \times 10^{-7} $   \\ \hline
mass set-2  & \bos $ 18.61 \times 10^{-7} $  & $ 17.98 \times 10^{-7} $
& $ 18.34 \times 10^{-7} $   \\ \hline
mass set-3  & \bos $ 18.44 \times 10^{-7} $  & $ 18.13  \times 10^{-7} $   
& $ 18.98 \times 10^{-7} $   \\ \hline
\end{tabular}
\vskip 0.3 cm
\caption{Branching ratios for the inclusive $B_d \rar K \tau^+ \tau^-$
decay in the 2HDM, with the long distance contributions.}
\end{center}                                     
\end{table}

\newpage


Finally in Table 5, we display the results for the ratio $R$ of the inclusive 
$B(b \rar s \tau^+ \tau^-)$ decay calculated in the 2HDM 
and SM value of the inclusive $B(b \rar s \tau^+ \tau^-)$ channel.


\begin{table}[h]
\begin{center}
\begin{tabular}{|c|c|c|c|}
\hline
      &\bos ~\,$ tan \beta = 1 $~\, & ~\,$ tan \beta = 20 $~\, & ~\,$
tan \beta = 30 $~\,
\\  \hline\hline
mass set-1  & \bos 7.27 & 6.91 & 7.09  \\ \hline
mass set-2  & \bos 7.16 & 6.92 & 7.05  \\ \hline
mass set-3  & \bos 7.09 & 6.97 & 7.30  \\ \hline
\end{tabular}
\vskip 0.3 cm
\caption{The ratio $R$ of the inclusive $B(b \rar s \tau^+ \tau^-)$ decay in
2HDM and SM.}
\end{center} 
\end{table}


In conclusion, we calculate the rare $B \rar K \ell^+ \ell^-$ decay in 
2HDM. It is observed that the forward-backward asymmetry $A_{FB}$, 
the longitudinal polarization $P_L$ and the transversal 
polarization $P_T$ of the charged final lepton are very sensitive to the
variations in $tan \beta$. Therefore, in search of new physics 
their experimental investigation can serve as the crucial test. 

%

\newpage
\section*{Figure Captions}
{\bf 1.} Invariant mass distribution for the 
$B \rar K \tau^+ \tau^-$ decay with the long distance effects.
In Fig.1 (a), (b) and (c) the numbers 1, 2 and 3 on each curve
identify the mass set-1, 2 and 3, respectively, for the Higgs particles. 
The curve numbered as 4 is the one calculated in the 
Standard Model. The sharp peaks in
these figures are all 
due to the long distance contributions. \\ \\
{\bf 2.} The dependence of the forward-backward asymmetry $A_{FB}$ on $p^2$
for the decay $B \rar K \tau^+ \tau^-$. The lines numbered as 1,3 and 5
represent the long distance
effects for the mass sets 1, 2 and 3, respectively, while the lines with 
numbers 2, 4 and 6 are
chosen for the short distance effects. \\ \\
{\bf 3.} The dependence of the longitudinal polarization, $P_L$, on
$p^2$ for the $B \rar K \tau^+ \tau^-$ without the long distance effects.
The lines numbered as 1, 2 and 3
represent the mass set-1, 2 and 3, respectively. The line with number 4 is
the one presented for the Standard Model calculations.\\ \\
{\bf 4.} Transversal polarization asymmetry, $P_T$, for the \tept 
decay as a function of $p^2$. In this set of figures the lines numbered 
as 1, 2 and 3
represent the mass set-1, 2 and 3, respectively. \\ \\
{\bf 5.} The same as in Fig.1, but for the 
         $B \rar K \mu^+ \mu^-$ decay. 
Fig.5 (a) is for mass set-1, (b) is for mass set-2 and (c) is presented for
mass set-3. 
The lines numbered as 1, 3 and 5 represent the long distance
effects for the mass sets 1, 2 and 3, respectively, while the numbers 2, 4 and 6 are
presented for the short distance effects. The abbreviation $SM$ on the
bottom pair of curves (5 and 6) stands for the Standard Model results. \\ \\
{\bf 6.} The same as in Fig.3 but for the \tepm decay.
The ordering of the figures and of the
numbering of lines are exactly the same as explained for Fig.5. \\ \\
{\bf 7.} The same as in Fig.4, but for the
         $B \rar K \mu^+ \mu^-$ decay.
Figures (a), (b) and (c) are presented for the mass set-1, mass set-2 and
mass set-3, respectively.

\begin{figure}
\vspace{25.cm}
    \includegraphics{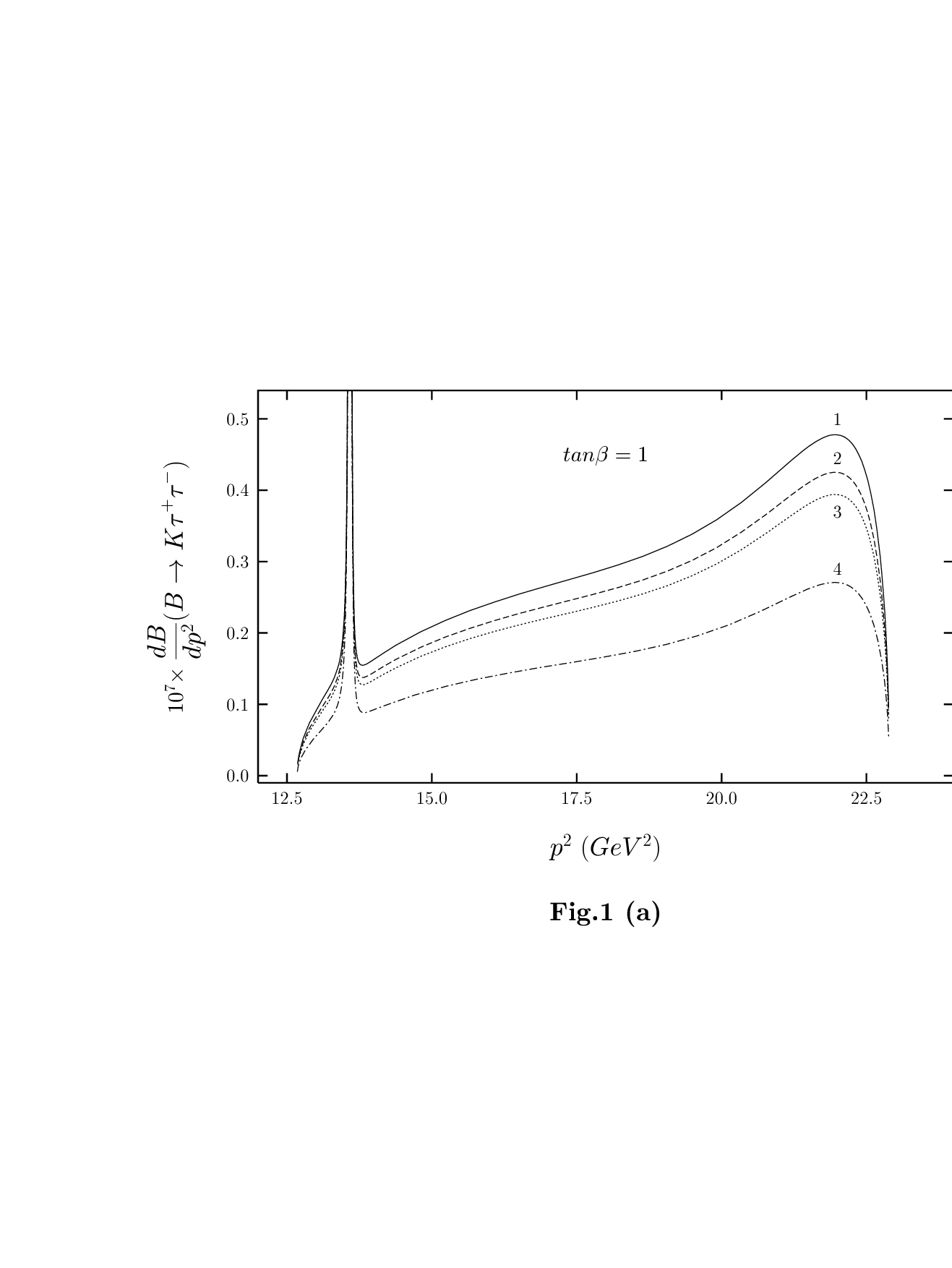}
    \includegraphics{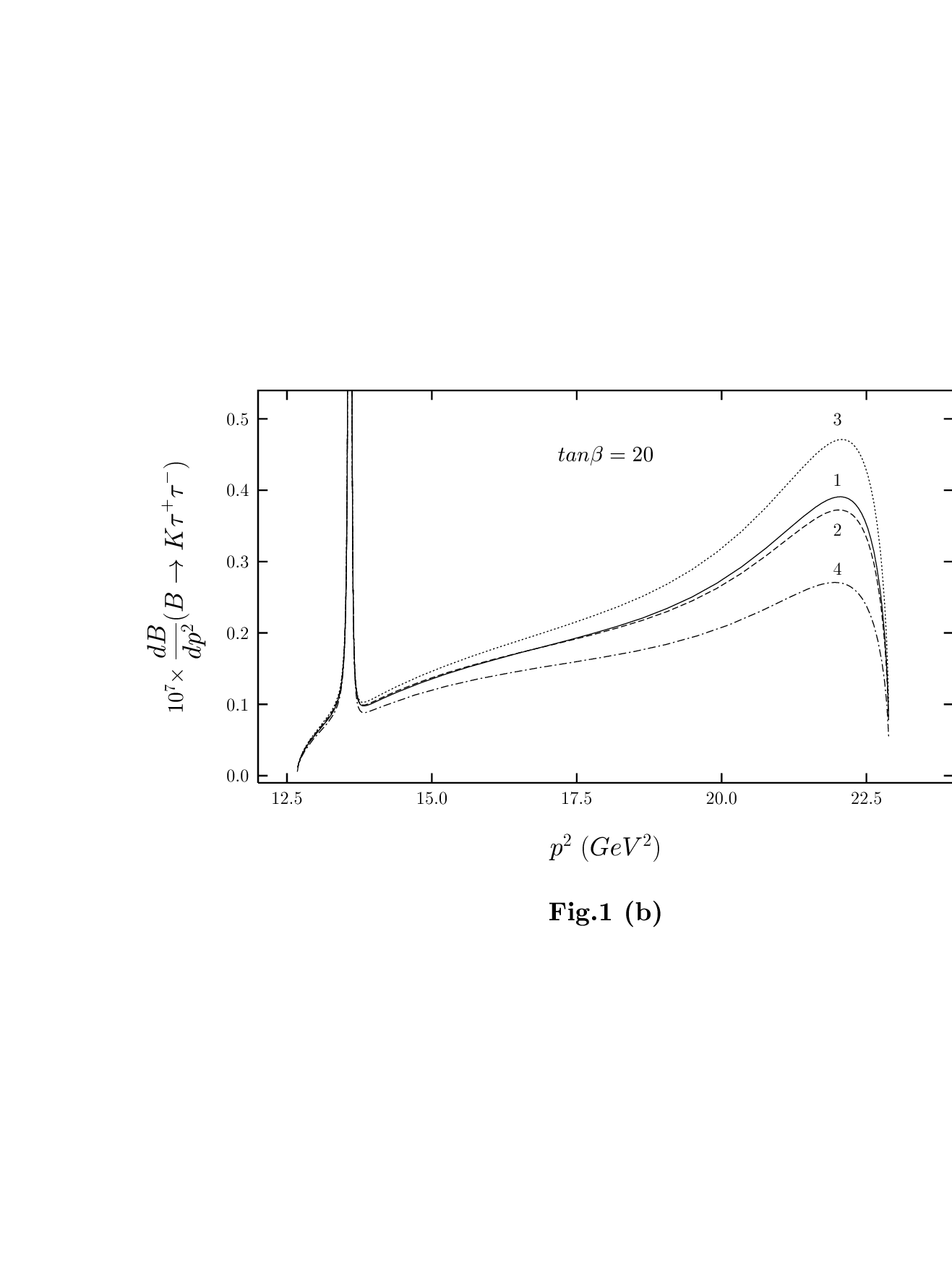}
    \vspace{-4.0cm}
\end{figure}

\begin{figure}
\vspace{25.cm}
    \includegraphics{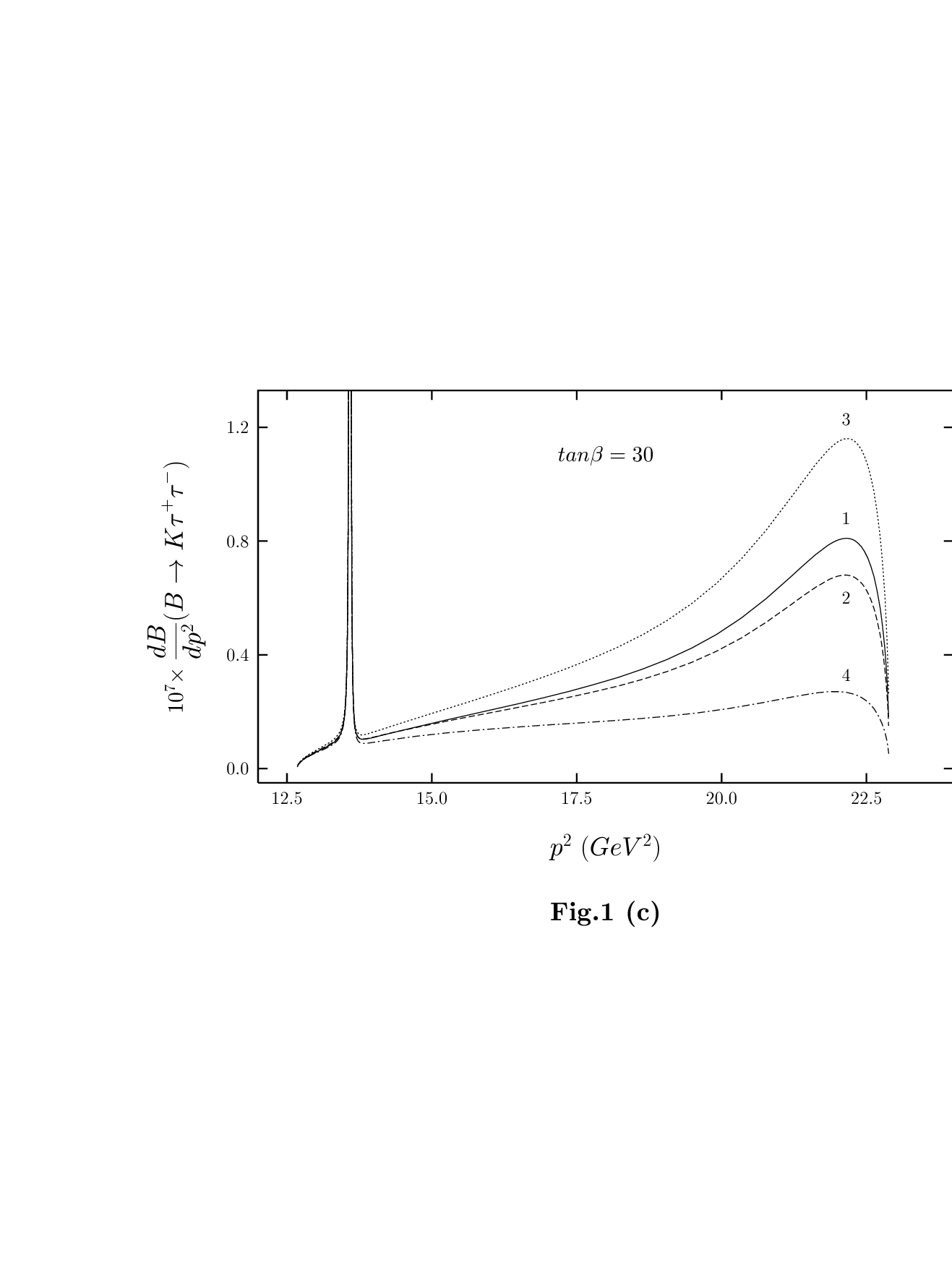}
    \includegraphics{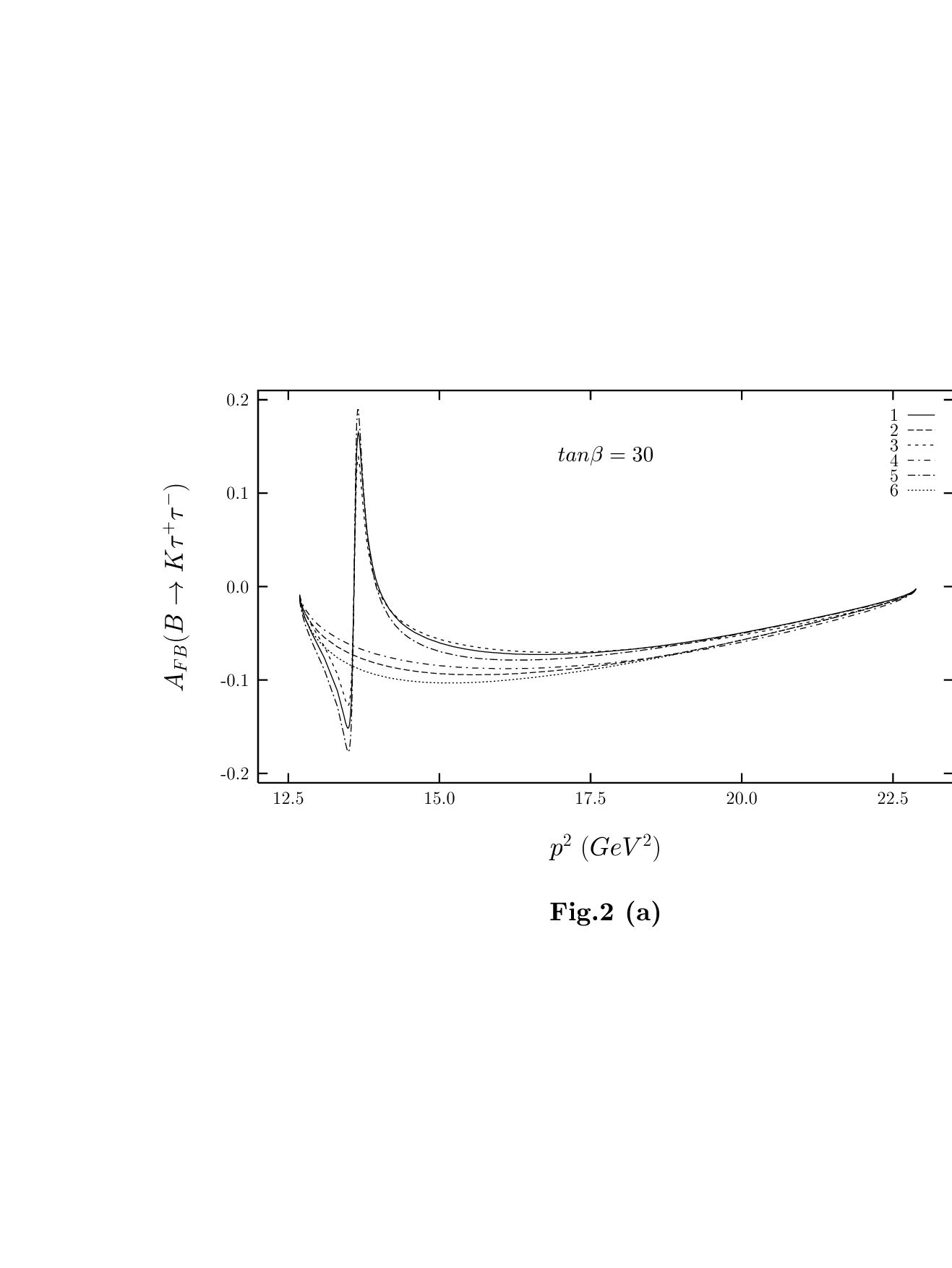}
    \vspace{-4.0cm}
\end{figure}

\begin{figure}
\vspace{25.cm}
    \includegraphics{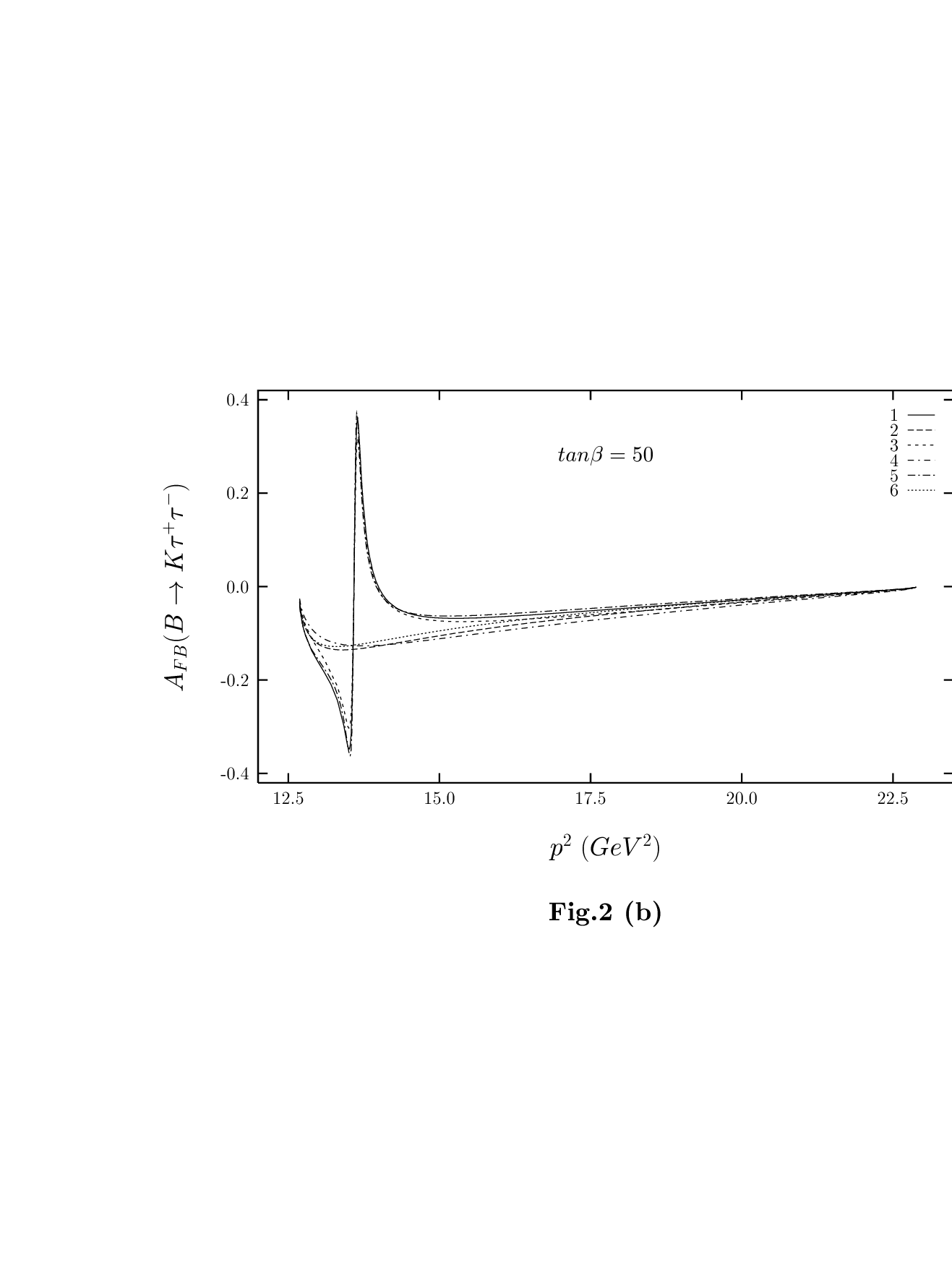}
    \includegraphics{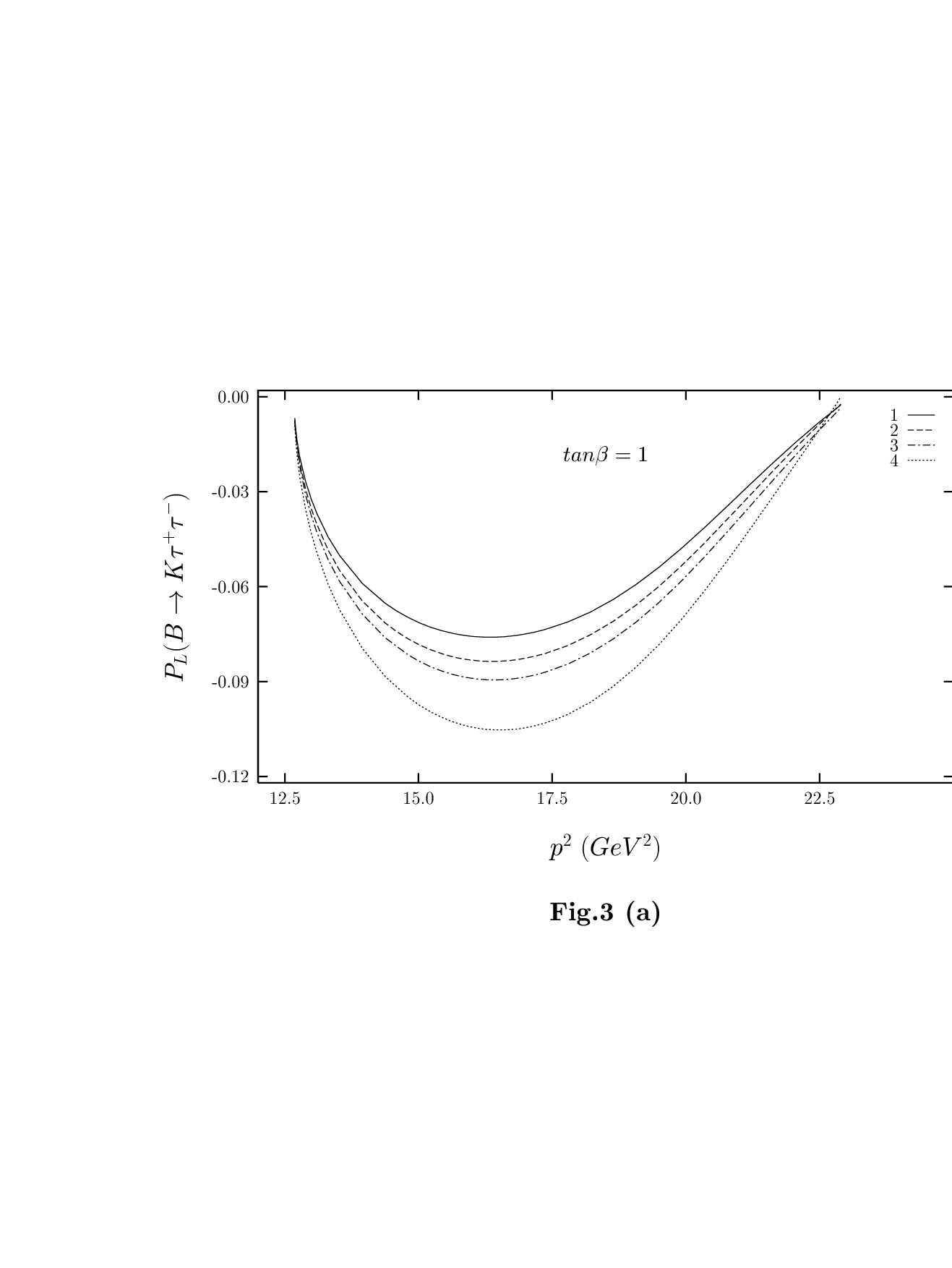}
    \vspace{-4.0cm}
\end{figure}  

\begin{figure}
\vspace{25.cm}
    \includegraphics{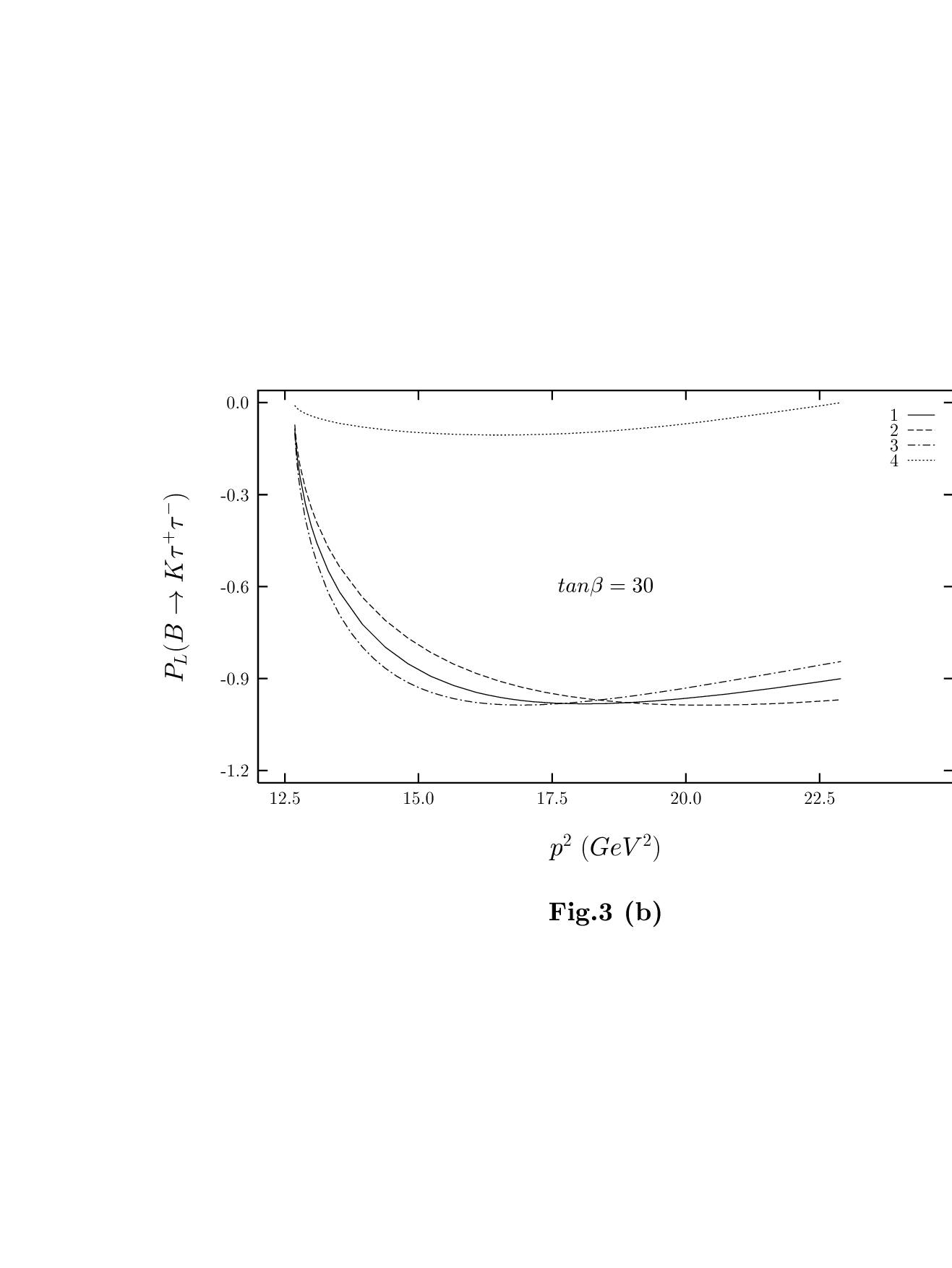}
    \includegraphics{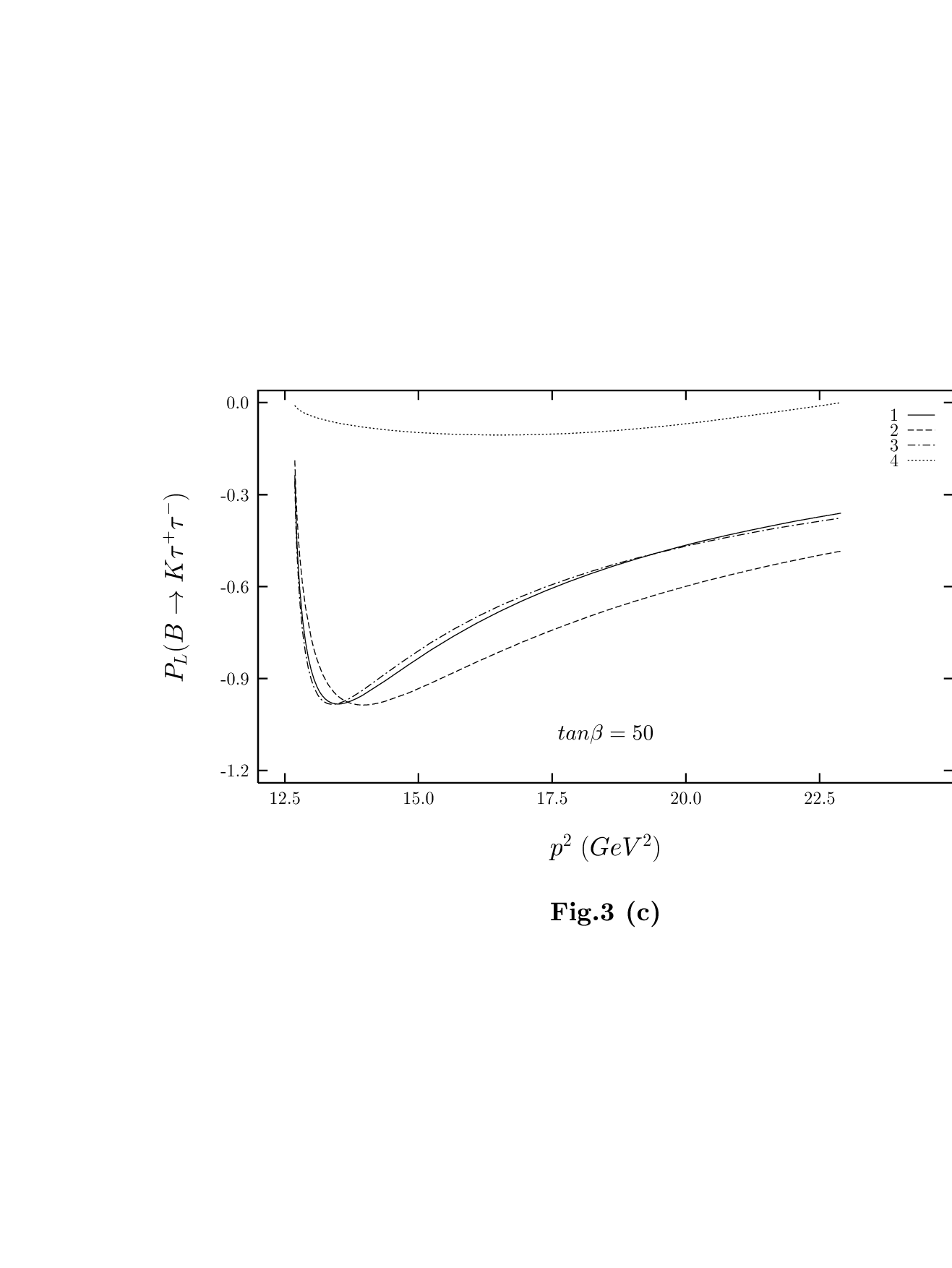}
    \vspace{-4.0cm}
\end{figure}

\begin{figure}
\vspace{25.cm}
    \includegraphics{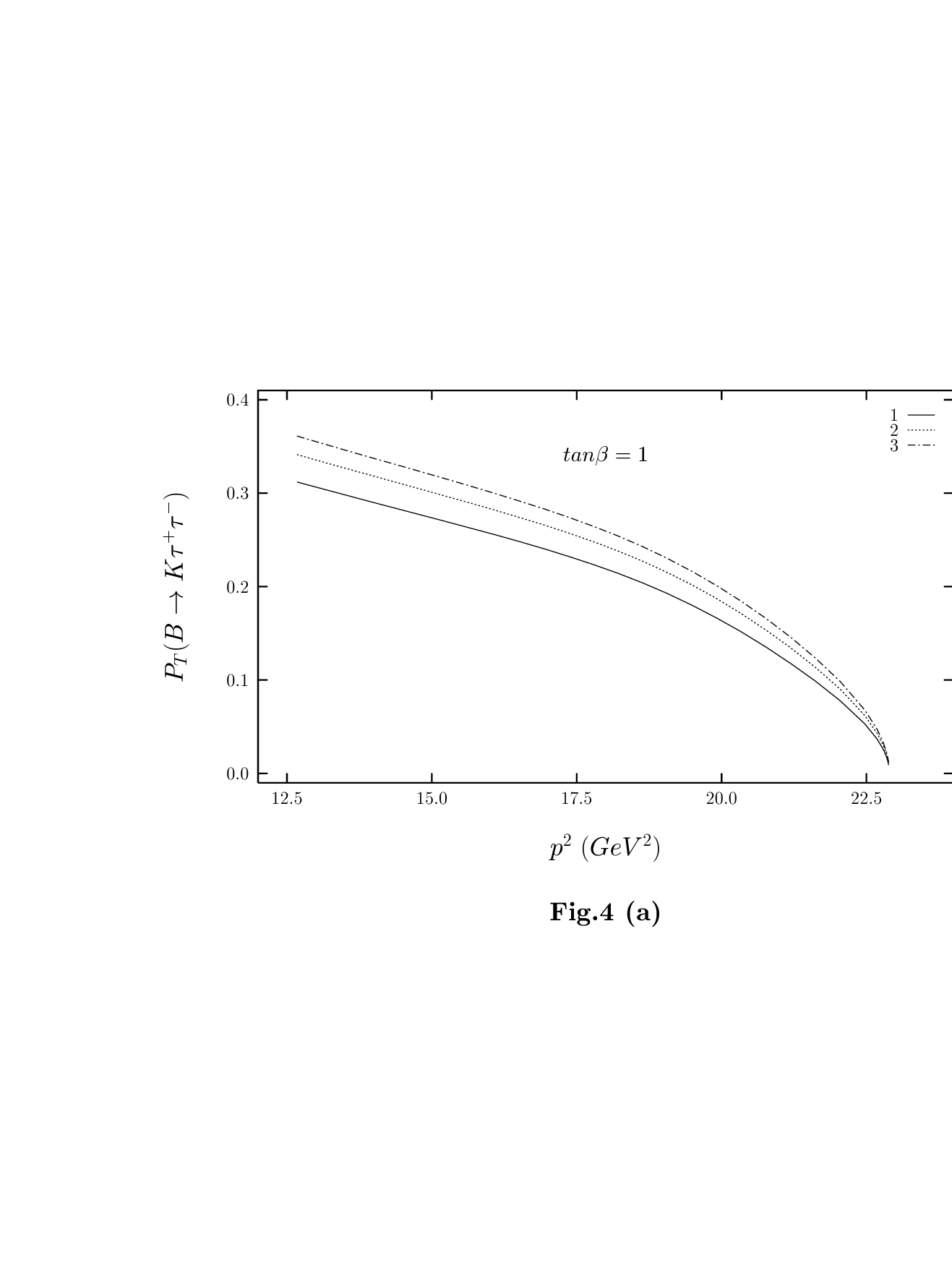}
    \includegraphics{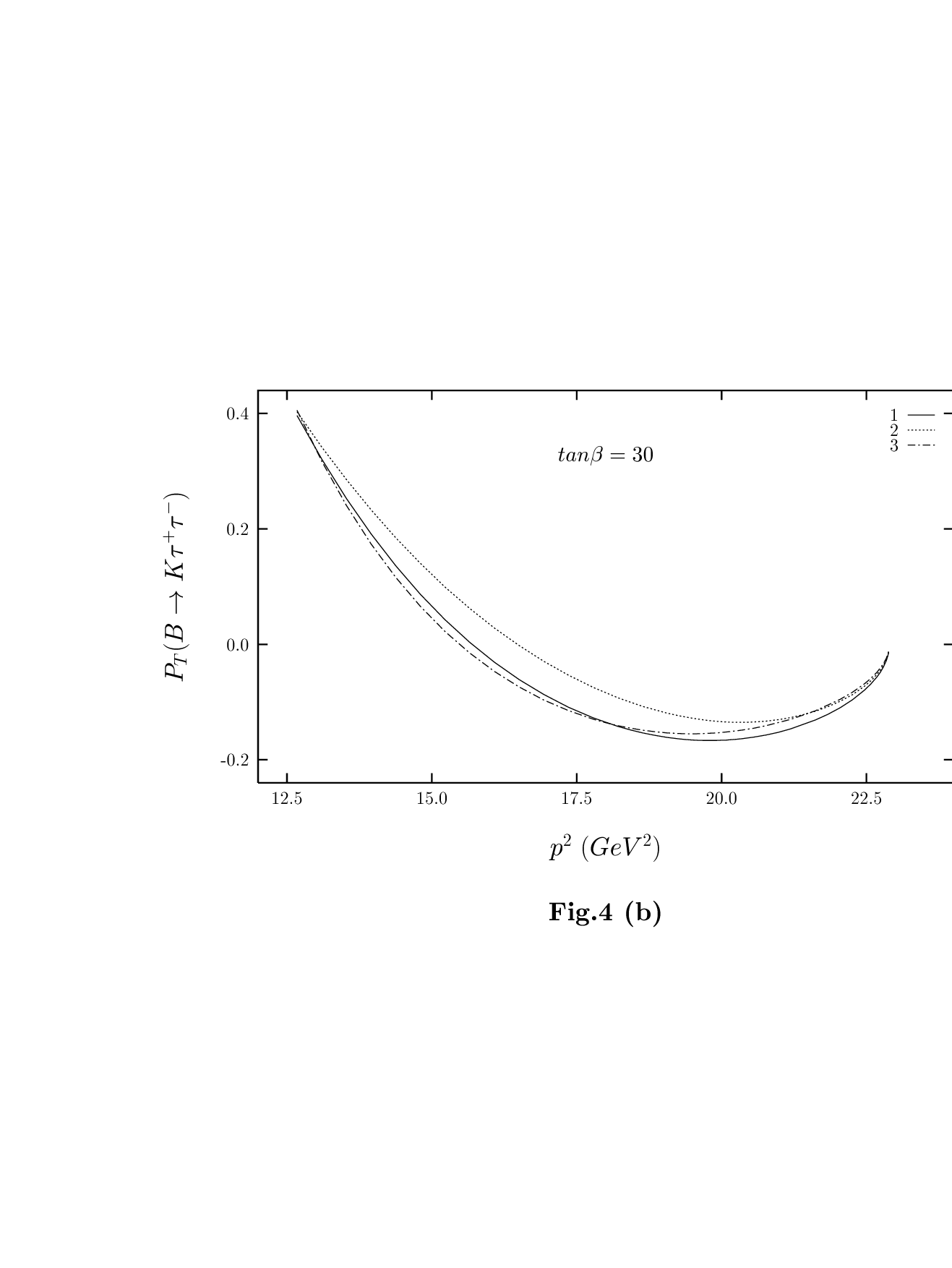}
    \vspace{-4.0cm}
\end{figure}

\begin{figure}
\vspace{25.cm}
    \includegraphics{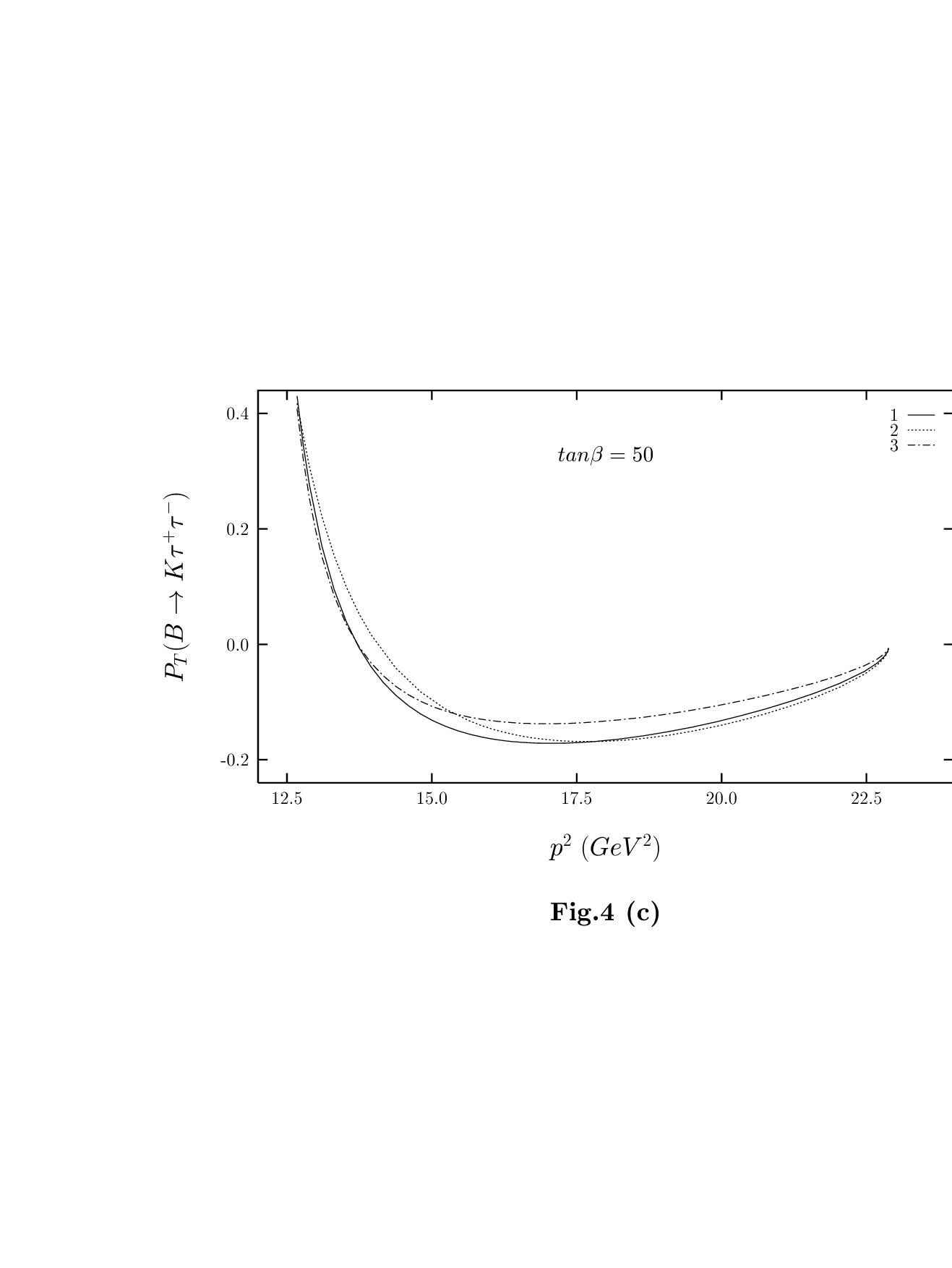}
    \includegraphics{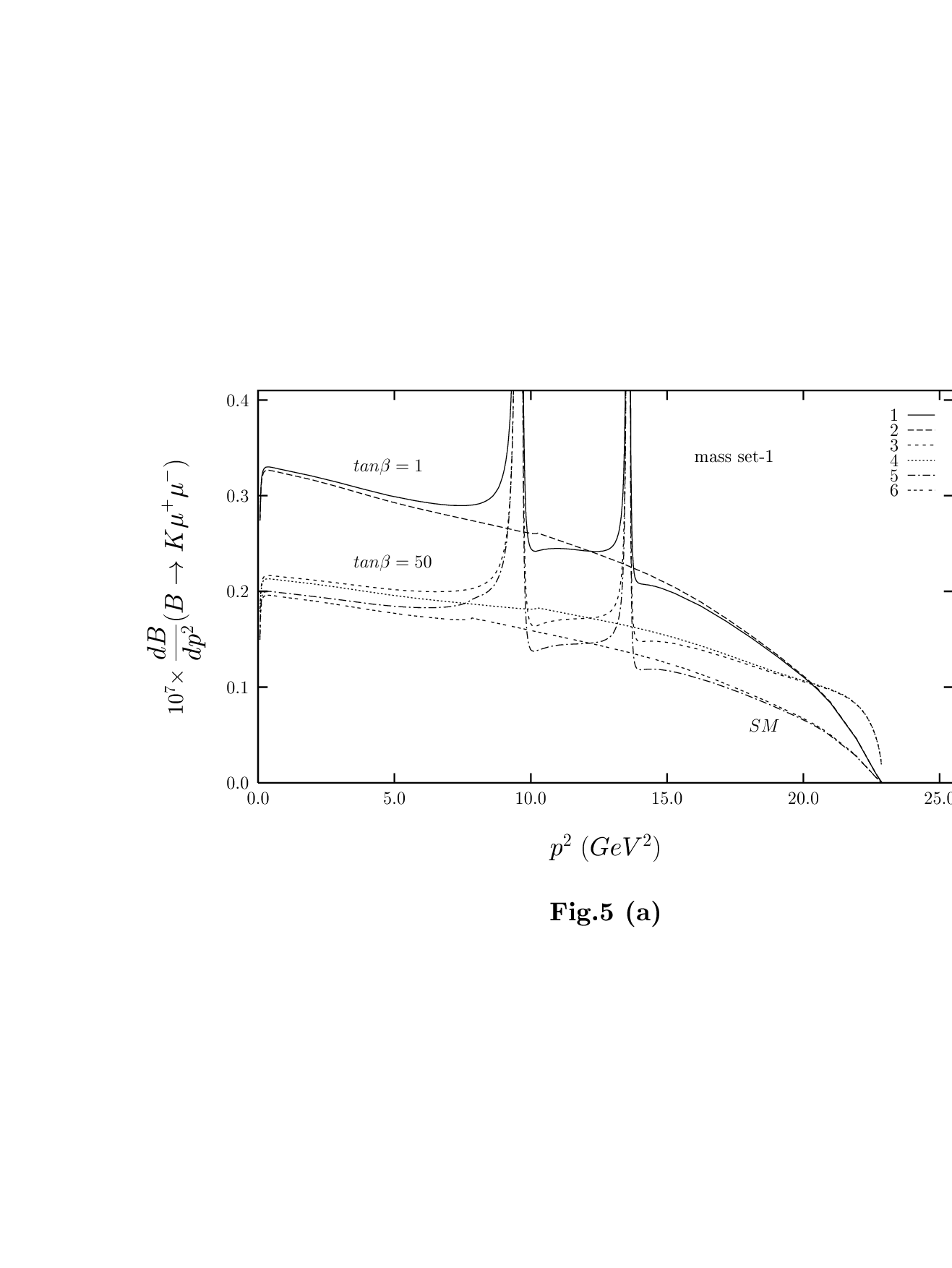}
    \vspace{-4.0cm}
\end{figure}

\begin{figure}
\vspace{25.cm}
    \includegraphics{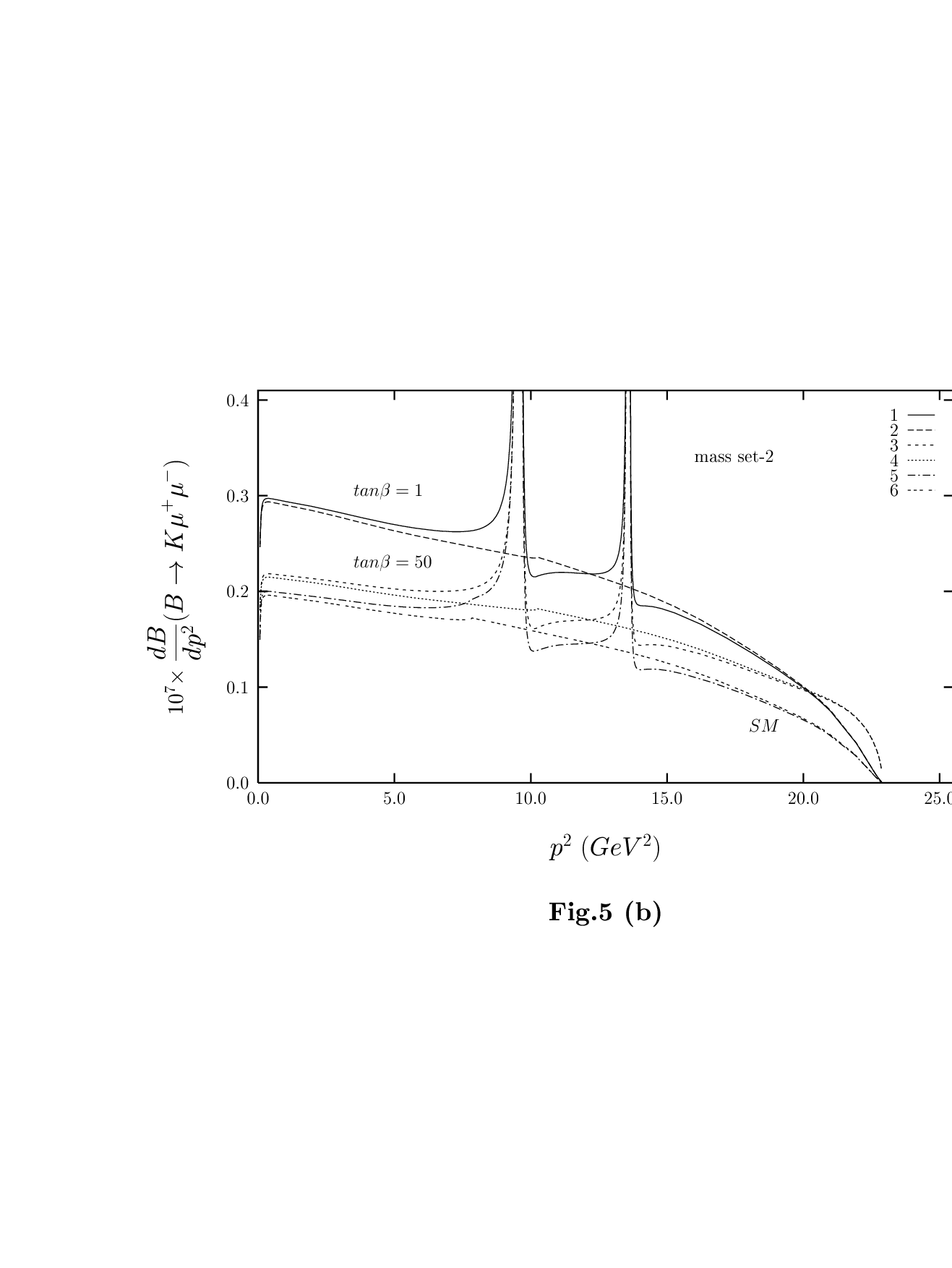}
    \includegraphics{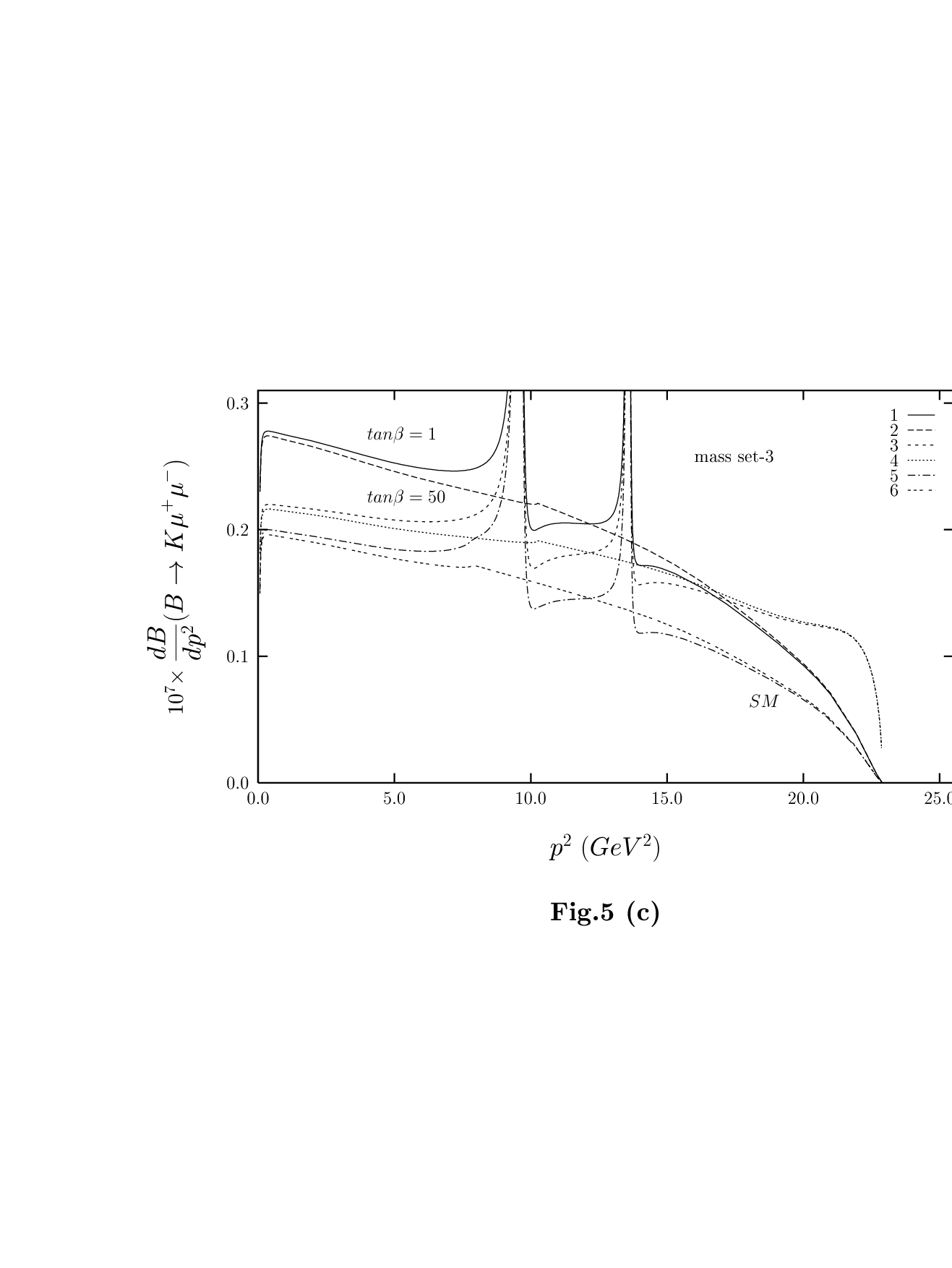}
    \vspace{-4.0cm}
\end{figure}

\begin{figure}
\vspace{25.cm}
    \includegraphics{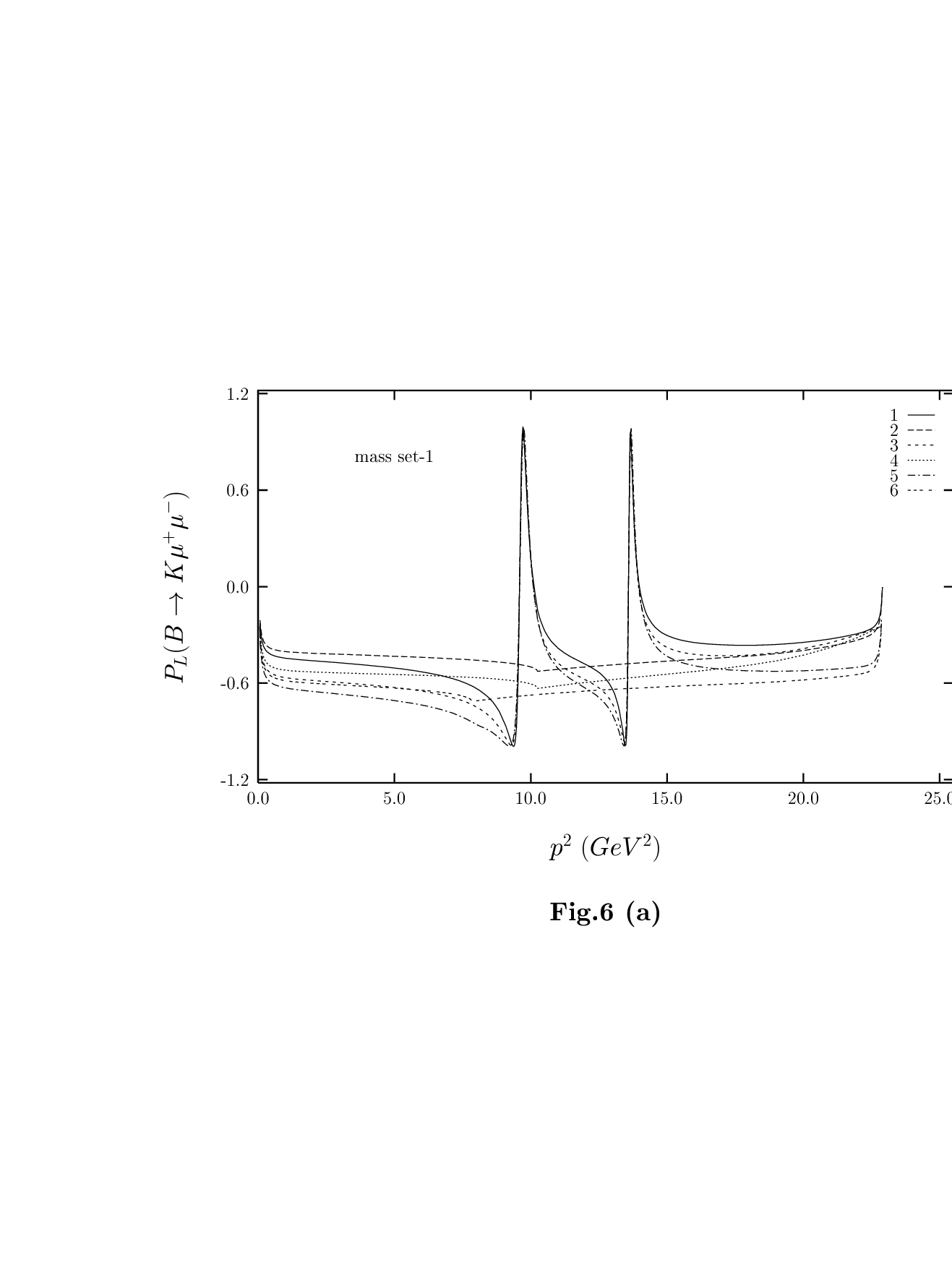}
    \includegraphics{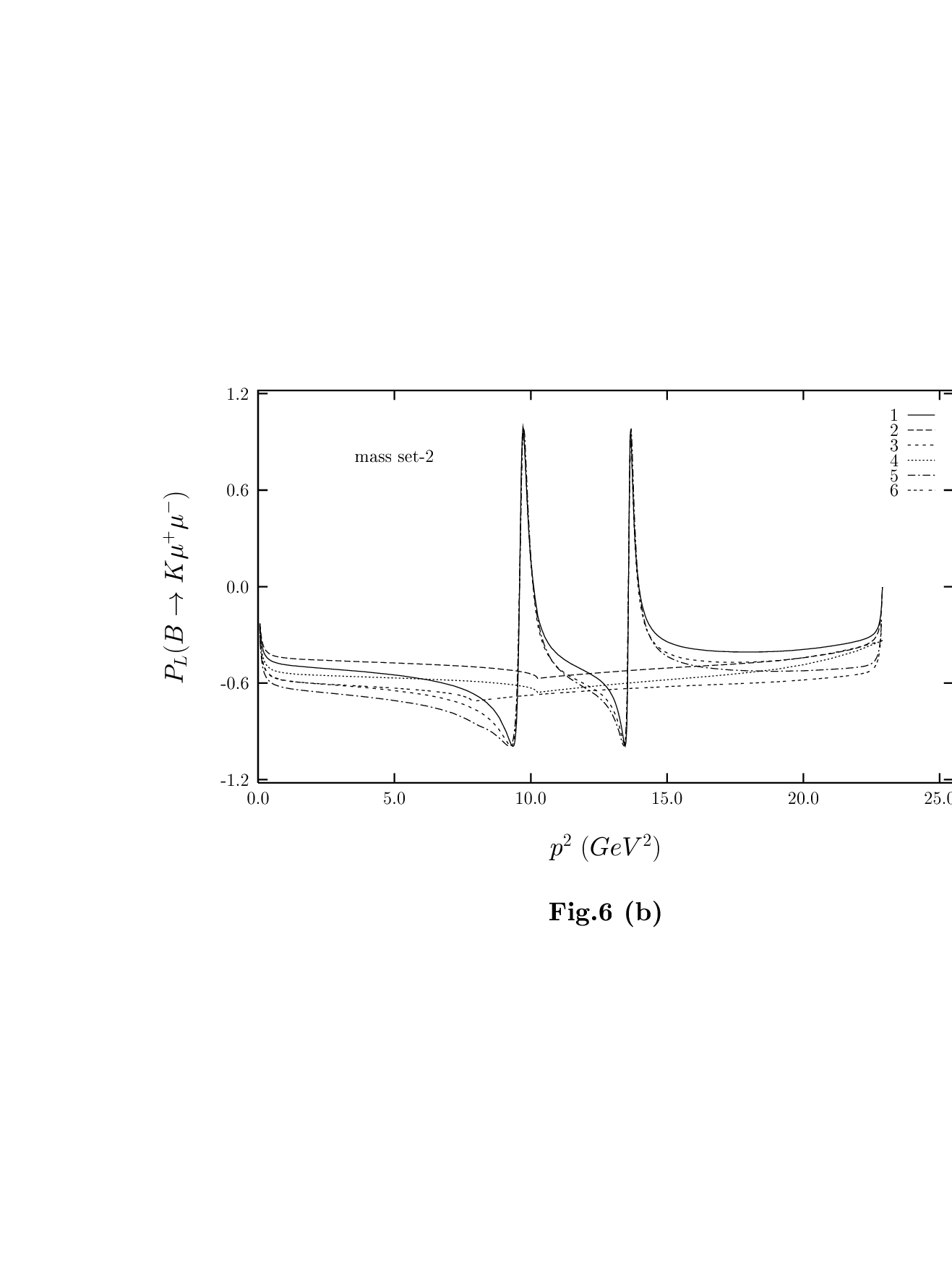}
    \vspace{-4.0cm}
\end{figure}

\begin{figure}
\vspace{25.cm}
    \includegraphics{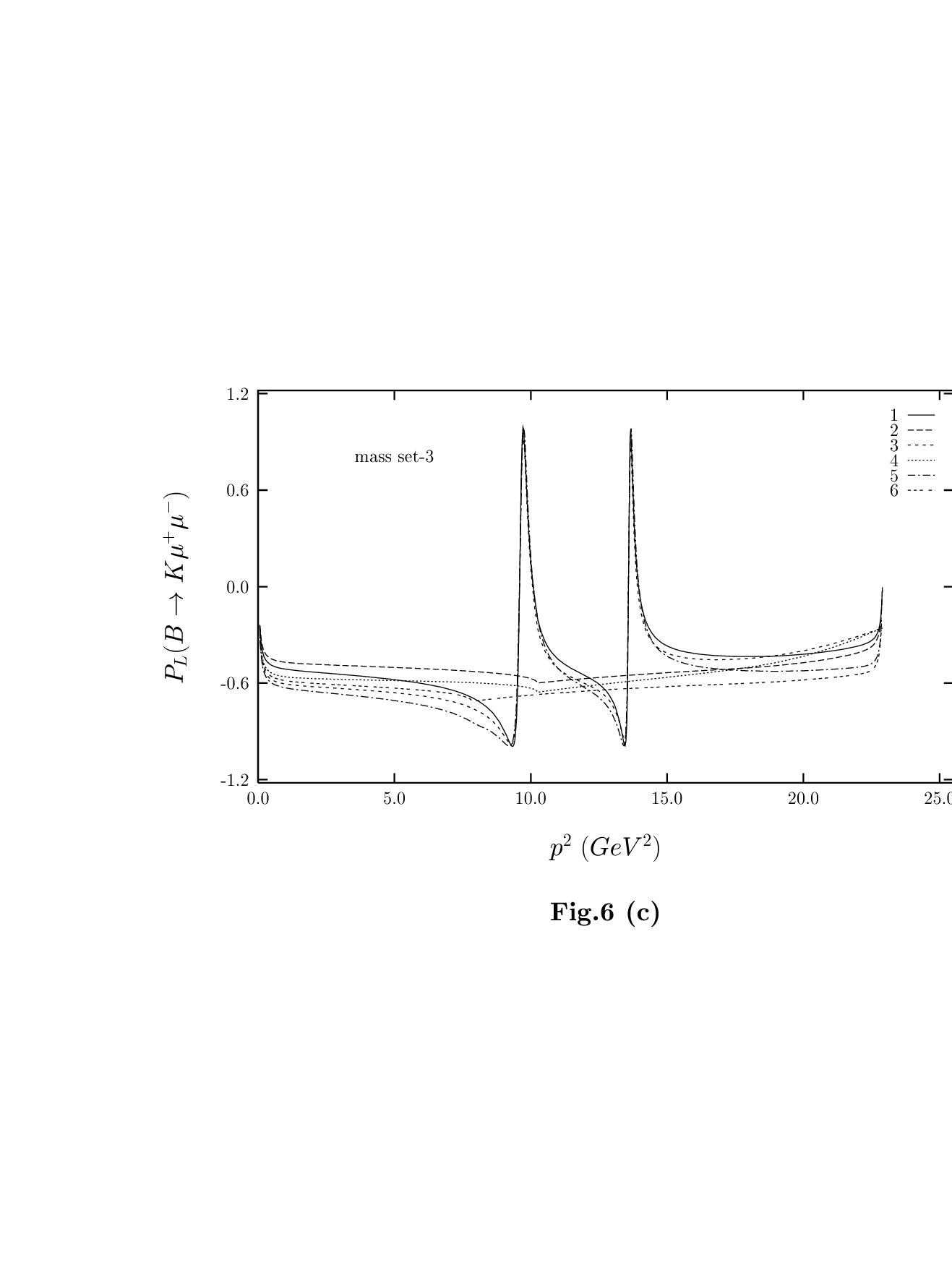}
    \includegraphics{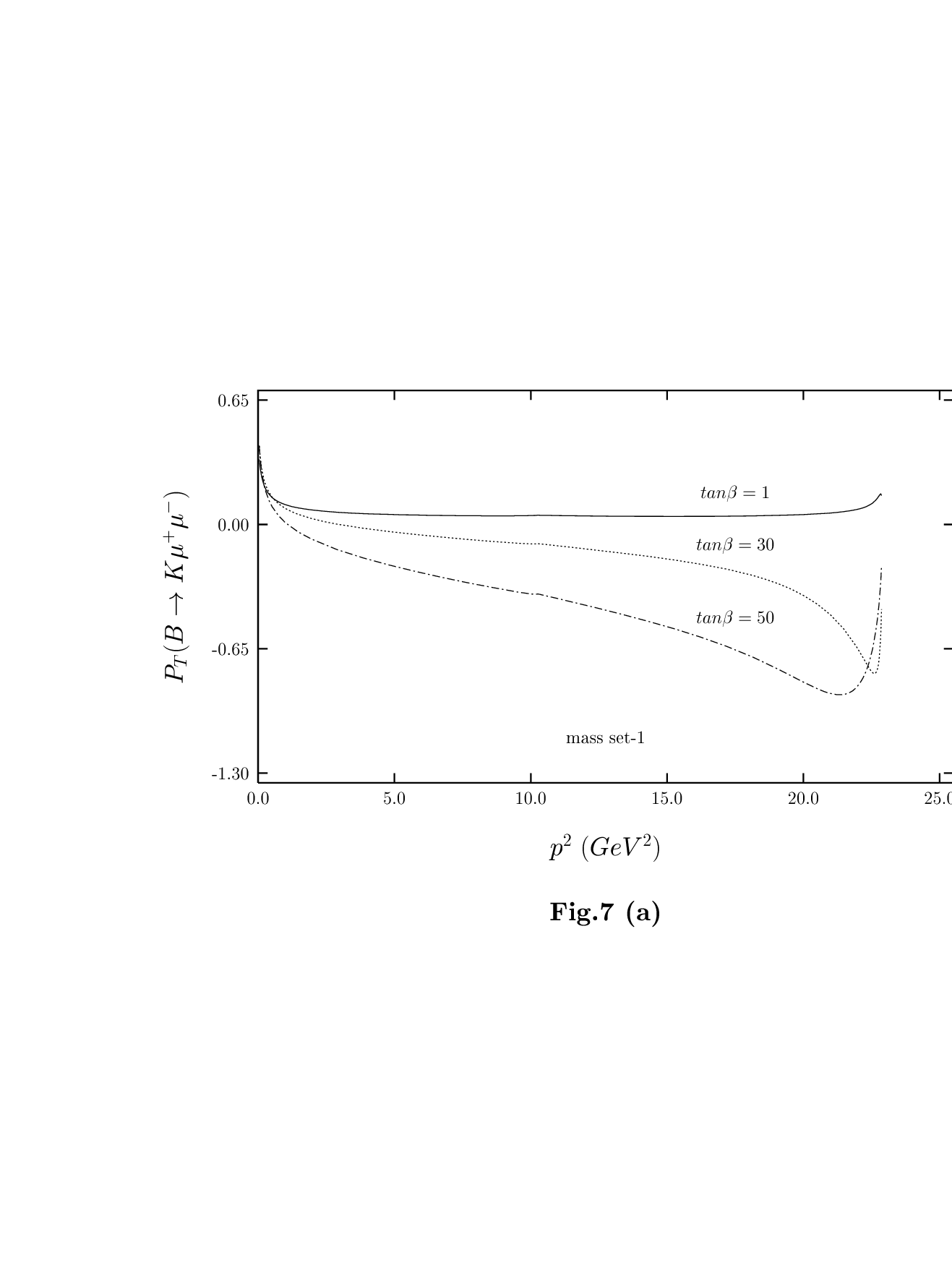}
    \vspace{-4.0cm}
\end{figure}

\begin{figure}
\vspace{25.cm}
    \includegraphics{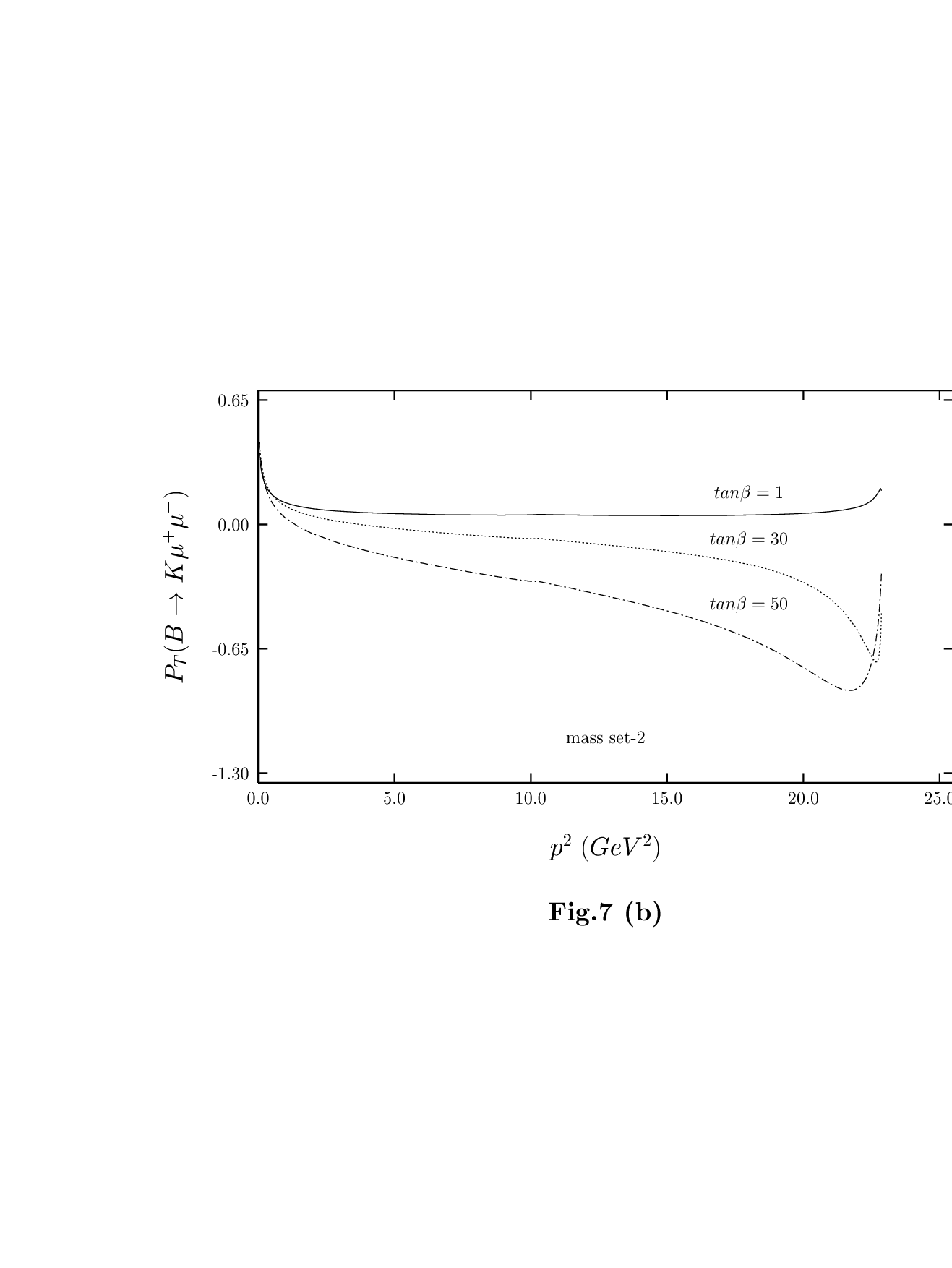}
    \includegraphics{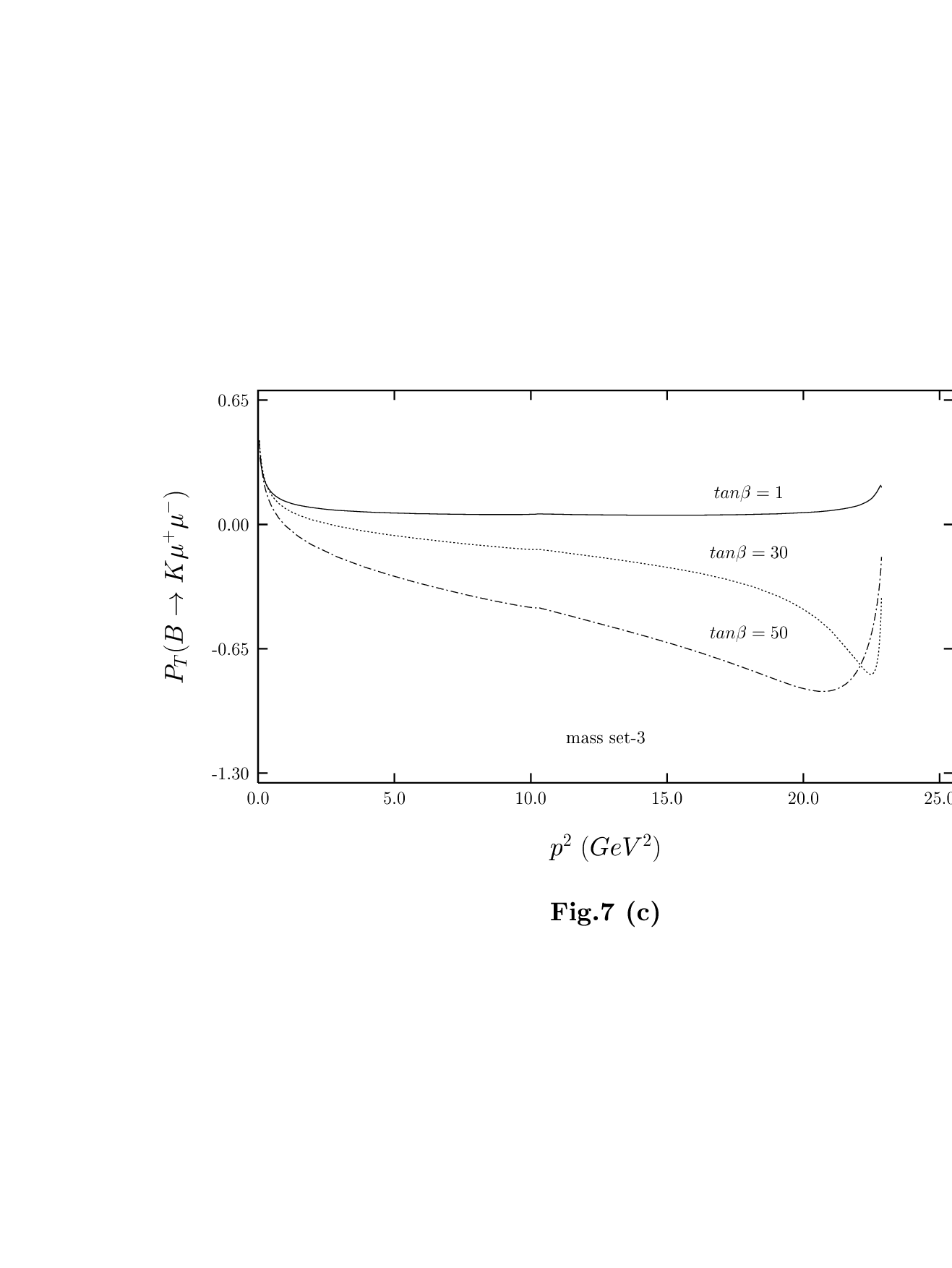}
    \vspace{-4.0cm}
                                \end{figure}  al


\newpage
\begin{thebibliography}{99}

\bibitem{R3} C. Anway-Wiese, {\bf CDF} Collaboration,
in: {\it Proc. of the} $8^{th}$ {\it Meeting of the Division of
Particle and Fields of the American Physical Society}, Albuquerque, 
New Mexico, 1994, ed: S. Seidel (World Scientific, Singapore, 1995).

\bibitem{R1} Z. Ligeti and M. Wise, {\it Phys. Rev.} {\bf D53} (1996) 4937;

\bibitem{R4} W. -S. Hou, R. S. Willey and A. Soni,
{\it Phys. Rev. Lett.} {\bf 58} (1987) 1608.

\bibitem{R5} N. G. Deshpande and J. Trampetic,
{\it Phys. Rev. Lett.} {\bf 60} (1988) 2583.

\bibitem{R6} C. S. Lim, T. Morozumi and A. I. Sanda,
{\it Phys. Lett.} {\bf B218} (1989) 343.

\bibitem{R7} B. Grinstein, M. J. Savage and M. B. Wise,
{\it Nucl. Phys.} {\bf B319} (1989) 271.

\bibitem{R8} C. Dominguez, N. Paver and Riazuddin, 
{\it Phys. Lett.} {\bf B214} (1988) 459.

\bibitem{R9} N. G. Deshpande, J. Trampetic and K. Ponose,
{\it Phys. Rev.} {\bf D39} (1989) 1461.

\bibitem{R10} W. Jaus and D. Wyler,
{\it Phys. Rev.} {\bf D41} (1990) 3405.

\bibitem{R11} P. J. O'Donnell and H. K. Tung,
{\it Phys. Rev.} {\bf D43} (1991) 2067.

\bibitem{R12} N. Paver and Riazuddin,
{\it Phys. Rev.} {\bf D45} (1992) 978.

\bibitem{R13} A. Ali, T. Mannel and T. Morozumi,
{\it Phys. Lett.} {\bf B273} (1991) 505.

\bibitem{R14} A. Ali, G. F. Giudice and T. Mannel,
{\it Z. Phys.} {\bf C67} (1995) 417.

\bibitem{R15} C. Greub, A. Ioannissian and D. Wyler,
{\it Phys. Lett.} {\bf B346} (1995) 145; \\
D. Liu {\it Phys. Lett.} {\bf B346} (1995) 355; \\
G. Burdman, {\it Phys. Rev.} {\bf D52} (1995) 6400: \\
Y. Okada, Y. Shimizu and M. Tanaka {\bf hep-ph}/9704223.

\bibitem{R16} A. J. Buras and M. M\"{u}nz,
{\it Phys. Rev.} {\bf D52} (1995) 186.

\bibitem{R17} N. G. Deshpande, X. -G. He and J. Trampetic,
{\it Phys. Lett.} {\bf B367} (1996) 362.

\bibitem{R2} J. F. Gunion, H. E. Haber, G.Kane and S. Dawson,
{\it "The Higgs Hunters Guide"}, (Addison-Wesley Reading, MA, 1990).

\bibitem{R22} A. K. Grant,
{\it Phys. Rev.} {\bf D51} (1995) 207.

\bibitem{R20} J. Kalinowski,
{\it Phys. Lett.} {\bf B245} (1990) 201. 

\bibitem{R18} J. L. Hewett, {\it Phys. Rev.} {\bf D53} (1996) 4964.

\bibitem{R19} F. Kr\"{u}ger and L. M. Sehgal,
{\it Phys. Lett.} {\bf B380} (1996) 199.

\bibitem{R23} R. Casalbuoni, A. Deandrea, N. Di Bartolomeo, 
R. Gatto and G. Nardulli,\\
{\it Phys. Lett.} {\bf B312} (1993) 315.

\bibitem{R24} P. Colangelo, F. De Fazio, P. Santorelli and E. Scrimieri,
{\it Phys. Rev.} {\bf D53} (1996) 3672.

\bibitem{R25} W. Jaus and D. Wyler,
{\it Phys. Rev.} {\bf D41} (1990) 3405.

\bibitem{R26} W. Roberts,
{\it Phys. Rev.} {\bf D54} (1996) 863.

\bibitem{R27} T. M. Aliev, H. Koru, A. \"{O}zpineci and M.Savc{\i},
{\it Phys. Lett.} {\bf B400} (1997) 194. 

\bibitem{R28} Yuan-Ben Dai, Chao-Shang Huang and Han-Wen Huang,\\
{\it Phys. Lett.} {\bf B390} (1997) 257.


\bibitem{R30} M. Misiak, 
{\it Nucl. Phys.} {\bf B398} (1993) 23;   
Erratum: {\it ibid} {\bf B439} (1995) 461.

\bibitem{R31} A. J. Buras and M. M\"{u}nz,
{\it Phys. Rev.} {\bf D52} (1995) 186;\\
M. Ciuchini, E. Franco, G. Martinelli,
L. Reina and L. Silvestrini, \\
{\it Phys. Lett.} {\bf B316} (1993) 127;\\
M. Ciuchini, E. Franco, G. Martinelli and L. Reina,
{\it Nucl. Phys.} {\bf B415} (1994) 403;\\
G. Cella, G. Curci, R. Ricciardi and A. Vicere,
{\it Nucl. Phys.} {\bf B421} (1994) 41; \\
{\it ibid} {\it Phys. Lett.} {\bf B325} (1994) 227.

\bibitem{R32} A. I. Vainshtein, V. I. Zakharov, L. B. Okun and M. A. Shifman,\\
{\it Sov. J. Nucl. Phys.} {\bf 24} (1976) 427.

\bibitem{R33} Particle Data Group,
{\it Phys. Rev.} {\bf D54} (1996).

\bibitem{R34} F. Kr\"{u}ger and L. M. Sehgal,
{\it Phys. Rev.} {\bf D55} (1997) 2799.

\bibitem{R35} A. Ali,
{\it Prep.} {\bf DESY} 97-019; {\bf hep-ph}/9702312 (1997).

\end{thebibliography}
\end{document}